\documentclass[10pt]{article} %***
\usepackage[sectionbib]{natbib}
\usepackage{array,epsfig,fancyheadings,rotating}
\usepackage[]{hyperref}  %<----modified by Ivan
\usepackage{sectsty, secdot}
\sectionfont{\fontsize{11}{12pt plus.8pt minus .6pt}\selectfont}
\renewcommand{\theequation}{\thesection\arabic{equation}}
\subsectionfont{\fontsize{11}{12pt plus.8pt minus .6pt}\selectfont}
\usepackage{amsmath}
\usepackage{amssymb}
\usepackage{amsfonts}
\usepackage{multirow}
\usepackage{amsthm}
\usepackage{comment}

\theoremstyle{definition}

\begin{document}

\renewcommand{\baselinestretch}{2}

\markright{ \hbox{\footnotesize\rm  
		%{\footnotesize\bf 24} (201?), 000-000
	}\hfill\\[-13pt]
	\hbox{\footnotesize\rm
		%\href{http://dx.doi.org/10.5705/ss.20??.???}{doi:http://dx.doi.org/10.5705/ss.20??.???}
	}\hfill }

\markboth{\hfill{\footnotesize\rm KHANDOKER AKIB MOHAMMAD AND OTHERS} \hfill}
{\hfill {\footnotesize\rm EFFICIENT ESTIMATION FOR COX PH CURE MODEL} \hfill}

\renewcommand{\thefootnote}{}
$\ $\par

%%%%%%%%%%%%%%%%%%%%%%%%%%%%%%%%%%%%%%%%%%%%%%%%%%%%%%%%%%%%%%%%%%%%%%%%%%%%%%%%%%%%%%%%%%%%%%%%%%%%%%%%%%%%%%%%%%%%%%%%%%%%

\fontsize{12}{14pt plus.8pt minus .6pt}\selectfont \vspace{0.8pc}
\centerline{\large\bf EFFICIENT ESTIMATION FOR THE  }
\vspace{2pt} \centerline{\large\bf COX PROPORTIONAL HAZARDS CURE MODEL}
\vspace{.4cm} \centerline{Khandoker Akib Mohammad$^\ast$, Yuichi Hirose, Budhi Surya and Yuan Yao} \vspace{.4cm} \centerline{\it
	Victoria University of Wellington} \vspace{.55cm} \fontsize{9}{11.5pt plus.8pt minus
	.6pt}\selectfont

%%%%%%%%%%%%%%%%%%%%%%%%%%%%%%%%%%%%%%%%%%%%%%%%%%%%%%%%%%%%%%%%%%%%%%%%%%%%%%%%%%%%%%%%%%%%%%%%%%%%%%%%%%%%%%%%%%%%%%%%%%%%

\begin{quotation}
	\noindent {\it Abstract:}
	%{\bf Contents of the Abstract.}\\
	While analysing time-to-event data, it is possible that a certain fraction of subjects will never experience the event of interest and they are said to be cured. When this feature of survival models is taken into account, the models are commonly referred to as cure models. In the presence of covariates,  the conditional survival function of the population can be modelled by using cure model which depends on the probability of being uncured (incidence) and the conditional survival function of the uncured subjects (latency), and a combination of logistic regression and Cox proportional hazards (PH) regression is used to model the incidence and latency respectively. 
	%In this paper, we take the profile likelihood approach to estimate the cumulative hazard and the regression parameters. 
	In this paper, we have shown the asymptotic normality of the profile likelihood estimator via asymptotic expansion of the profile likelihood and obtain the explicit form of the variance estimator with an implicit function in the profile likelihood. We have also shown the efficient score function based on projection theory and the profile likelihood score function are equal. Our contribution in this paper is that we have expressed the efficient information matrix as the variance of the profile likelihood score function.
	%Moreover, the estimators of the regression parameters from the Cox PH cure model are shown to be semiparametric efficient. 
	A simulation study suggests that the estimated standard errors from bootstrap samples (SMCURE package) and the profile likelihood score function (our approach) are providing similar and comparable results.
	%can overcome by using efficient score function analytically (our approach). 
	The numerical result of our proposed method is also shown by using the melanoma data from SMCURE R-package \citep{Cai et al.} and we compare the results with the output obtained from SMCURE package.

	\vspace{9pt}
	\par
\end{quotation}\par

\def\thefigure{\arabic{figure}}
\def\thetable{\arabic{table}}

\renewcommand{\theequation}{\thesection.\arabic{equation}}

\fontsize{12}{14pt plus.8pt minus .6pt}\selectfont

\setcounter{section}{1} %***
\setcounter{equation}{0} %-1

\lhead[\footnotesize\thepage\fancyplain{}\leftmark]{}\rhead[]{\fancyplain{}\rightmark\footnotesize\thepage}%Put this line in Page 2

\noindent {\bf 1. Introduction}\\
In survival analysis, Cox PH cure model has attracted attention for decades. \cite{Kuk and Chen} first proposed the Cox PH cure model as a semiparametric generalization of Farewell's model (1982) where a combination of Cox PH model and logistic regression has been used to study the survival times of uncured subjects and cure rate respectively. In clinical settings, Cox PH cure model has been widely used for modelling the failure time data for various types of cancer studies such as breast cancer, head and neck cancer,  leukemia, prostate cancer, melanoma etc \citep{Peng and Dear, Sy and Taylor, Sy and Taylor (a), Zhao and Zhou, Othus et al., Peng and Taylor, Amico and Keilegom}.

The efficiency and asymptotic distribution of semiparametric maximum likelihood estimator have been studied for the Cox PH cure model by \cite{Fang}. Later a non-parametric maximum likelihood approach has been used to find the estimator of the cumulative hazard and the regression parameters from the Cox PH cure model, and the asymptotic properties are established by the modern empirical process theory \citep{Lu}. The joint maximization approach developed by \cite{Murphy, Murphy (a)} has been used by \cite{Lu} to find the efficient estimators for Cox PH cure model. However, the above works of efficiency and asymptotic distribution of maximum likelihood estimator  did not address the computation with the implicit function in the profile likelihood estimation. Later, \cite{Cai et al.} developed an R package (SMCURE) to fit the Cox PH cure model which have received much attention in recent years \citep{Peng and Taylor, Amico and Keilegom}. In SMCURE package, \cite{Cai et al.} used the melanoma data from the ECOG phase III clinical trial e1684, where the standard errors of the estimated parameters  have been calculated by using bootstrap methods.

\cite{Hsieh et al.} pointed out that in some examples, the estimator of baseline hazard function based on the profile likelihood approach is an implicit function.
For these examples, including the model in this paper, it was very challenging to show asymptotic normality of the profile likelihood estimator. For this reason, \cite{Hsieh et al.} proposed to use Bootstrap method \citep{Efron and Tibshirani} to get the standard errors while using the profile likelihood approach.

In the papers (\citealp{Zeng and Cai,Zeng and Lin,Zeng and Lin (a)}), the baseline hazard and regression parameters have been maximized jointly where the idea of equality between the profile likelihood estimator and the maximum likelihood estimator has been used. Maximizing the profile likelihood with respect to regression parameters lead us to same estimate as one obtained by maximizing the likelihood jointly with respect to regression parameters and baseline hazard. In these papers, the asymptotic distributions of the estimated regression parameters and baseline hazard have shown by jointly maximizing the likelihood function where projection theory has been used to compute the efficient score function.  Using these results, the asymptotic normality of the estimated regression parameters has been shown without going through profile likelihood expansion. 

\cite{Murphy and Vaart} have invented a version of profile likelihood approach in semiparametric models as an inferential tool.  They have used an 'approximate least favorable submodel' to deal with implicit function under the profile likelihood approach.  Their proposed approach has been used to differentiate the 'approximate least favorable submodel' and didn't involve the differentiation of the profile likelihood function. 

All of these works (\citealp{Murphy and Vaart,Zeng and Cai,Zeng and Lin,Zeng and Lin (a)}) have used the techniques which can avoid differentiation of the implicit function under the profile likelihood function and therefore they didn't derive the score function based on the profile likelihood. Ultimately, they have shown the asymptotic variance of the estimator as the  inverse of the efficient information matrix. However, they have not shown the efficient information matrix in terms of the profile likelihood score function. 

In this paper, we profile out the baseline hazard function from Cox PH cure model and plugged the estimator in the likelihood function. However the problem is that the estimator of the baseline hazard function is an implicit function \citep{Rizopoulos}. We solve the difficulty of showing asymptotic normality of the estimator (Theorem-3 and Theorem-4 in Section 3.3). This approach is alternative to the methodologies where the asymptotic normality of profile likelihood estimator has been studied (\citealp{Murphy and Vaart,Zeng and Cai,Zeng and Lin,Zeng and Lin (a),Hirose (b),Hirose-c}). In this paper, we have used the asymptotic expansion of profile likelihood function to get the asymptotic normality of the profile likelihood estimator and obtain the explicit form of the variance estimator using the profile likelihood score function. These results can be used in computation to calculate the estimated variance of the profile likelihood estimator which is illustrated in the simulation study (Section-4) and in numerical example (Section-5). For the numerical example, we have used the data (ECOG phase III clinical trial e1684) from SMCURE package and computed the standard errors of the estimated parameters  from the efficient information matrix (based on the profile likelihood score function). 
%We have also shown that the efficient score function based on projection theory (the ordinary score function of regression parameters minus its orthogonal projection onto the closed linear span of the score functions for the baseline hazard) and the profile likelihood score function (based on the differentiation of the profile likelihood function) are equal.

Our contribution in this paper is that we have expressed the efficient information matrix as the variance of the score function in the profile likelihood. This gives us not only analytical understanding of the profile likelihood estimation, but also numerical method to compute the efficient information matrix using the profile likelihood score function. 

%By using the asymptotic expansion of the likelihood, we have found the explicit form of the efficient score function and established the asymptotic normality of the profile likelihood estimator. Hence we got the explicit form of the estimate of variance for the profile likelihood estimator. 
This paper is organized as follows. A brief discussion on Cox PH cure model has been given in Section-2. In Section-3, we describe the estimation procedure and theorems which are used to show that the profile likelihood estimators are consistent and asymptotically normal. Results obtained from the profile expansion of Cox PH cure model are shown in Section-4 and Section-5. This paper concludes in Section-6  with a short discussion.

\setcounter{equation}{0} %-1
%\noindent {\bf 2. The Second Section}
\section{Cox PH Cure Model}
%{\bf Contents of the second section.}
Let us define a binary variable $V$, where $V=0$ indicates an individual that will be a long-term survivor (never experience the event of interest) and $V=1$ indicates an individual that will experience the event. For an individual with covariate vector $W=(1, W_1,..., W_n)$, the distribution of $V=1$ can be expressed as a logistic model 
\begin{equation} \label{2.1}
p=Pr(V=1, W;b)=\frac{e^{b'W}}{1+e^{b'W}},
\end{equation}
where $p$ is the probability of being susceptible (often called incidence of the model), $b$ is a vector parameter and $W$ include the intercept. The time to experience the event among individuals for which $V=1$ can be modelled by Cox PH model
\begin{equation} \label{2.2}
\lambda(t|V=1,Z; \beta)=\lambda_0(t|V=1)e^{\beta'Z},
\end{equation}
where we observe another set of covariate $Z$ without intercept and $\lambda_0(t|V=1)$ is the baseline hazard function. The two sets of covariates may be identical, or partially or completely different from each other \citep{Kuk and Chen}.

An individual who experience the event at time $t$ contributes a likelihood factor
\begin{equation*}
p f(t|V=1,Z; \lambda, \beta),
\end{equation*}
which is the probability of death at time $t$ \citep{Kuk and Chen}. On the other hand, an individual who has been followed to time $t$ without experiencing the event contributes a likelihood factor
\begin{equation*}
(1-p)+pS(t|V=1,Z; \lambda, \beta),
\end{equation*}
which is the probability of long-term survivor (cure) plus the probability of experiencing the event after time $t$ \citep{Kuk and Chen}. In addition $S(t|V=1,Z; \lambda, \beta)=S_0(t|V=1)e^{\beta'Z}$ is the conditional survival function of the susceptibles (often called the latency) where $S_0(t|V=1)=\exp \big(-\Lambda_0(t|V=1) \big)=\exp \big(-\int_{0}^{t}\lambda_0(s|V=1)ds \big)$ is the baseline survival function and $\Lambda_0(t|V=1)$ is the baseline cumulative hazard function.

\setcounter{equation}{0} %-1
%\noindent {\bf 3. The Third Section}
\section{Estimation}
Suppose the observed data for individual $i$ can be denoted by $(T_i,\delta_i, Z_i); i= 1, 2...,n$ where $T_i$ is the length of time a subject was observed, $Z_i$ is a vector of covariates. Moreover, $\delta_i$ indicates whether the observed time is censored or not
$$\delta_i=
\left\{ \begin{array}{rcl}
1 & \mbox{for}
& T_i=\textrm{event time} \\ 0 & \mbox{for} & T_i=\textrm{censored time}  
\end{array}\right.$$

For convenience, let $W_i=(1, Z'_i)'$, although the covariates in $W_i$ and $Z_i$ do not have to be equal.

The likelihood for $n$ observations will be
\begin{equation} \label{ 3.1}
L(b, \beta, \lambda)=\prod_{i=1}^{n} \bigg\{p_i f(t_i|V=1,Z_i; \lambda, \beta)\bigg\}^{\delta_i}  \bigg\{(1-p_i)+p_iS(t_i|V=1,Z_i; \lambda, \beta)\bigg\}^{1-\delta_i},
\end{equation}
where $p_i$ is the probability of $i$th individual being susceptible. We know that $$f(t|V=1,Z; \lambda, \beta)=\lambda(t|V=1,Z; \beta)S(t|V=1,Z; \lambda, \beta).$$

So for the Cox PH cure model, the observed full likelihood function can be written as
\begin{equation} \label{ 3.2}
\begin{aligned}
L(b, \beta, \Lambda_0)={} & \prod_{i=1}^{n}\bigg[p_i \lambda(t_i|V=1,Z_i; \beta)S(t_i|V=1,Z_i; \lambda, \beta)\bigg]^{\delta_i}\bigg[(1-p_i)+p_iS(t_i|V=1,Z_i; \lambda, \beta) \bigg]^{1-\delta_i}.
%& \bigg[(1-p_i)+p_iS(t_i|V=1,Z_i; \lambda, \beta) \bigg]^{1-\delta_i}.
\end{aligned}
\end{equation}

Here we want to obtain the estimates of $b$ and $\beta$ that maximize $L(b, \beta, \Lambda_0)$. 
%That is why we cannot use ordinary Cox PH model because the partial likelihood does not depend on $\lambda_0(t)$. 
For maximizing $L(b, \beta, \Lambda_0)$, we are going to apply profile likelihood technique in which $\Lambda_0(t)$ is profiled out from the likelihood.\\
\textbf{3.1 The Expectation-Maximization (EM) Algorithm}

Let us define the complete data by $(t_i,\delta_i, Z_i, v_i), ~i=1,..., n$ which includes the observed data and unobserved $v_i$, where $v_i$ is the value taken by the variable $V_i$. It follows that if $\delta_i=1$ then $v_i=1$ and if $\delta_i=0$ then $v_i$ is unobserved. The choice for using EM algorithm is justified by the fact that the model depends on a latent variable, $v_i$ (cure status). Moreover, the aim of 
EM algorithm is to maximize observed data likelihood from a complete data likelihood \citep{Dempster}. So the complete data likelihood can be written as
\begin{equation} \label{3.3}
\begin{aligned}
L_c(b, \beta, \Lambda_0;v)={} & \prod_{i=1}^{n}\bigg[p_i \lambda(t_i|V=1,Z_i; \beta) S(t_i|V=1,Z_i; \lambda, \beta)\bigg]^{\delta_i v_i}\\
&\times \prod_{i=1}^{n}\bigg[p_i S(t_i|V=1,Z_i; \lambda, \beta)\bigg]^{(1-\delta_i) v_i}
\times \prod_{i=1}^{n} \bigg[1-p_i\bigg]^{(1-\delta_i)(1-v_i)}.
\end{aligned}
\end{equation}

The above equation can be rewritten as the product of a logistic and a PH component.
\begin{equation} \label{3.4}
\begin{aligned}
L_c(b, \beta, \Lambda_0;v)={} & \prod_{i=1}^{n}p_i^{v_i}(1-p_i)^{1-v_i} \times \prod_{i=1}^{n}\lambda(t_i|V=1,Z_i; \beta)^{\delta_i v_i}S(t_i|V=1,Z_i; \lambda, \beta)^{v_i}.
%={} &  L_1(b ;v). L_2(\beta, \Lambda_0;v)%=\prod_{i=1}^{n}L_1(b ;v)L_2(\beta, \lambda;v)
\end{aligned}
\end{equation}

So it is possible to estimate the incidence and the latency separately \citep{Amico and Keilegom}. Now the expected complete data log-likelihood under $p(V|T, \delta, Z)$ is
\begin{equation} \label{3.5}
\small
\begin{aligned}
%\sum_{V}^{}p(V|T, \delta, Z) \log L_c(b, \beta, \lambda;v) & ={} 
\sum_{i=1}^{n} \bigg\{\gamma(V_i) \log p_i+(1-\gamma(V_i)) \log (1-p_i) \bigg\} + \sum_{i=1}^{n}\gamma(V_i) \bigg\{\delta_i \log \lambda(t_i|V=1,Z_i; \beta)+ \log S(t_i|V=1,Z_i; \lambda, \beta)\bigg\},
\end{aligned}
\end{equation}
where $\gamma(V_i)$ can be defined as
\begin{equation} \label{3.6}
\begin{aligned}
\gamma(V_i)={} & E(V_i|T_i, \delta_i, Z_i)=\bigg(\frac{p_i S(t_i|V=1,Z_i; \lambda, \beta)}{1-p_i+p_i S(t_i|V=1,Z_i; \lambda, \beta)} \bigg)^{1-\delta_i}.%\\
%={} & p(V_i=1|T_i, \delta_i, Z_i)\\
%={} &\frac{p_i \lambda(t_i|V=1,Z_i; \beta)^{\delta_i}S(t_i|V=1,Z_i; \lambda, \beta)}{\big(p_i \lambda(t_i|V=1,Z_i; \beta)S(t_i|V=1,Z_i; \lambda, \beta)\big)^{\delta_i}\big(1-p_i+p_i S(t_i|V=1,Z_i; \lambda, \beta)\big)^{1-\delta_i}}\\
%={} & \bigg(\frac{p_i S(t_i|V=1,Z_i; \lambda, \beta)}{1-p_i+p_i S(t_i|V=1,Z_i; \lambda, \beta)} \bigg)^{1-\delta_i}.
\end{aligned}
\end{equation}

Here, for censored cases $\gamma(V_i)=E(V_i|T_i, \delta_i, Z_i)$ and for uncensored cases $\gamma(V_i)=1$. To estimate all parameters and the baseline hazards simultaneously, we combine the EM algorithm and profile likelihood approach. From equation (3.4), it can be observed that the likelihood function for the logistic component is same as for a classical logistic regression model. To estimate the parameters for incidence, we can apply the Newton-Raphson technique.\\
\textbf{Baseline Hazard Estimation}

Before starting the EM algorithm, we profile out the baseline hazard function $\lambda_0(t)$ using NPMLE (non-parametric maximum likelihood estimation). The survival part of equation (3.5) can be separately maximized with respect to $\lambda$ using the log-likelihood:
\begin{equation} \label{3.7}
%\sum_{i=1}^{n} \gamma(V_i) \log L_2(\beta, \lambda;v)=
\sum_{i=1}^{n} \gamma(V_i)\bigg[\delta_i\big\{\log \lambda_i+\beta' Z_i\big\}-e^{\beta' Z_i}\sum_{j=1}^{n}\lambda_j1\{t_j\leq t_i\}\bigg].
\end{equation}

Now from the derivative with respect to $\lambda_k$, we get
$$\hat{\lambda}_k(t|V=1; \beta)=\frac{\delta_k}{\sum_{l=1}^{n} \gamma(V_l) 1\{t_k\leq t_l\}e^{\beta' Z_l}}.$$%=\frac{\delta_k}{\sum_{l=1}^{n} \gamma(V_l) Y_l(t_k)e^{\beta' Z_l}}.$$

%Denote $\hat{\lambda}(\beta)= \big(\hat{\lambda}_1(\beta),..., \hat{\lambda}_n( \beta) \big)$.
So the estimate of the baseline cumulative hazard, $\Lambda(t)$ will be
\begin{equation} \label{3.8}
\hat{\Lambda}(t|V=1;\beta)=\sum_{i=1}^{n}\frac{\delta_i1\{t_i\leq t\}}{\sum_{l=1}^{n} \gamma(V_l) 1\{t\leq t_l\}e^{\beta' Z_l}}.%=\sum_{i=1}^{n}\hat{\lambda}_i(t|V=1;b, \beta)1\{t_i\leq t\}.
\end{equation}
\textbf{The E-step}

In the E-step, we use the current parameter estimates $b$ and $\beta$ to find the expected values of $V_i$:
\begin{equation} \label{3.9}
\gamma(V_i)= E(V_i|T_i, \delta_i, Z_i)=\bigg(\frac{p_i S \big(t_i|V=1,Z_i; \hat{\lambda}(\beta), \beta \big)}{1-p_i+p_i S \big(t_i|V=1,Z_i; \hat{\lambda}(\beta), \beta \big)} \bigg)^{1-\delta_i}.
\end{equation}
\textbf{The M-step}

By replacing $\lambda$ with $\hat{\lambda}(\beta)$, we maximize the equation (3.5)
\begin{equation} \label{3.10}
\small
\sum_{i=1}^{n} \bigg[ \bigg\{\gamma(V_i) \log p_i+(1-\gamma(V_i)) \log (1-p_i) \bigg\}
+ \gamma(V_i) \bigg\{\delta_i \log \hat{\lambda}(t_i|V=1,Z_i; \beta)+ \log S(t_i|V=1,Z_i; \hat{\lambda}(\beta), \beta)\bigg\} \bigg],
\end{equation}
with respect to $b$ and $\beta$ to obtain $\hat{b}$ and $\hat{\beta}$ respectively. The estimated parameters from the M-step are returned into E-step until the values of $\hat{b}$ and $\hat{\beta}$ converge.\\
\textbf{3.2 Score Functions}

An estimator of the baseline cumulative hazard function in the counting process notation \citep{Fleming} can be written from equation (3.8) as
\begin{equation} \label{3.11}
\hat{\Lambda}(t)=\int_{0}^{t}\frac{\sum_{i=1}^{n} dN_i(u)}{\sum_{i=1}^{n} \gamma(V_i) Y_i(u)e^{\beta' Z_i}},
\end{equation}
where $N(t)=1\{T\leq t,\delta=1\}$ and $Y (t)=1\{T \geq t\}.$

Let us denote $E_{F_n} f= \int fdF_n$. Then $\hat{\Lambda}(t)$ can be expressed as
\begin{equation} \label{3.12}
\hat{\Lambda}_{\beta, F_n}(t)=\int_{0}^{t}\frac{E_{F_n} dN(u)}{E_{F_n} \gamma(V) Y(u)e^{\beta' Z}}.
\end{equation}

Now from (3.10), the log-profile likelihood can be written as
\begin{equation} \label{3.13}
%\sum_{V}^{}p(V|T, \delta, Z) \log L_c(b, \beta, \hat{\Lambda}_{\beta, F_n};v)=
\sum_{i=1}^{n} \bigg\{\log P(V_i|b)+\log P \big(T_i, \delta_i|\hat{\Lambda}_{\beta, F_n}, \beta \big) \bigg\},%\\
%+ \sum_{i=1}^{n}\gamma(V_i) \big\{\delta_i \log \lambda(t_i|V=1,Z_i; \beta)+ \log S(t_i|V=1,Z_i; \hat{\Lambda}(\theta, F_n), \beta)\big\}.
\end{equation}
where $\log P(V_i|b)$ and $\log P \big(T_i, \delta_i|\hat{\Lambda}_{\beta, F_n}, \beta \big)$ are the log-profile likelihood functions (for one observation) for logistic and Cox PH component respectively. Now we can express the components as
\begin{equation} \label{3.14}
\begin{aligned}
\log P(V_i|b)={} & \big\{\gamma(V_i) \log p_i+(1-\gamma(V_i)) \log (1-p_i) \big\}\\%={} & \gamma(V_i) \log \frac{p_i}{1-p_i}+\log (1-p_i)\\
={} & \gamma(V_i)b'W_i- \log (1+ e^{b'W_i}),
\end{aligned}
\end{equation}
and
\begin{equation} \label{3.15}
\begin{aligned}
\log P \big(T_i, \delta_i|\hat{\Lambda}_{\beta, F_n}, \beta \big)={} & \gamma(V_i) \bigg\{\delta_i \log \hat{\lambda}(t_i|V=1,Z_i; \beta)+ \log S \big(t_i|V=1,Z_i; \hat{\Lambda}_{\beta, F_n}, \beta \big)\bigg\}\\
={} & \gamma(V_i)\bigg[\delta_i\big\{\log \frac{E_{F_n} dN(T_i)}{E_{F_n} \gamma(V) Y(T_i)e^{\beta' Z}}+\beta' Z_i\big\}-e^{\beta' Z_i}\int_{0}^{T_i}\frac{E_{F_n} dN(u)}{E_{F_n} \gamma(V) Y(u)e^{\beta' Z}}\bigg].
\end{aligned}
\end{equation}

The score functions for the profile likelihood are
\begin{equation} \label{3.16}
\phi(V_i, T_i, \delta_i|b, \beta, F_n)=\phi_l(V_i|b)+\phi_s(T_i, \delta_i|\beta, F_n),
\end{equation}
where $\phi_l(V_i|b)$ is the score function for logistic component which can be expressed as
\begin{equation} \label{3.17}
\begin{aligned}
\phi_l(V_i|b)={} & \frac{\partial}{\partial b}\log P(V_i|b)=\gamma(V_i)W_i- \frac{W_ie^{b'W_i}}{1+ e^{b'W_i}},
%\\
%={} & \gamma(V_i) \frac{\partial}{\partial \theta} b'W_i- \frac{\partial}{\partial \theta} \log (1+ e^{b'W_i})\\
%={} & \gamma(V_i)W_i- \frac{W_ie^{b'W_i}}{1+ e^{b'W_i}},
\end{aligned}
\end{equation}
and $\phi_s(T_i, \delta_i|\beta, F_n)$ is the score function for survival component which can be written as
\begin{equation} \label{3.18}
\small
\begin{aligned}
\phi_s(T_i, \delta_i|\beta, F_n)={} & \frac{\partial}{\partial \beta} \log P(T_i, \delta_i|\hat{\Lambda}_{\beta, F_n}, \beta)\\
%={} & \gamma(V_i) \frac{\partial}{\partial \beta} \bigg[\delta_i\big\{\log \frac{E_{F_n} dN(T_i)}{E_{F_n} \gamma(V) Y(T_i)e^{\beta' Z}}+\beta' Z_i\big\}-e^{\beta' Z_i}\int_{0}^{T_i}\frac{E_{F_n} dN(u)}{E_{F_n} \gamma(V) Y(u)e^{\beta' Z}}\bigg]\\
={} & \gamma(V_i) \bigg\{\delta_i \bigg[Z_i-\frac{E_{F_n} \gamma(V) Y(T_i)Ze^{\beta' Z}}{E_{F_n} \gamma(V) Y(T_i)e^{\beta' Z}} \bigg]-e^{\beta' Z_i} \int_{0}^{T_i}  \bigg[Z_i-\frac{E_{F_n} \gamma(V) Y(u)Ze^{\beta' Z}}{E_{F_n} \gamma(V) Y(u)e^{\beta' Z}} \bigg] d\hat{\Lambda}_{\beta, F_n}(u)    \bigg\}.
\end{aligned}
\end{equation}

Now we will calculate the score function $B(T_i, \delta_i|\beta,F)$, which is Hadamard differentiable with respect to $F$. For an integrable function $h$ with the same domain as $F$, we can express
\begin{equation*}
\begin{aligned}
B(T_i, \delta_i|\beta, F)h={} &  d_{F} \log P(T_i, \delta_i|\beta, \hat{\Lambda}_{\beta, F})h\\
={} & \gamma(V_i) \bigg\{\delta_i \bigg[\frac{E_{h} dN(T_i)}{E_{F} dN(T_i)}  -\frac{E_{h} \gamma(V) Y(T_i)e^{\beta' Z}}{E_{F} \gamma(V) Y(T_i)e^{\beta' Z}} \bigg]
-e^{\beta' Z_i} \int_{0}^{T_i} \frac{E_{h} dN(u)}{E_{F} \gamma(V) Y(u)e^{\beta' Z}}\\
& +e^{\beta' Z_i} \int_{0}^{T_i} \frac{E_{F} dN(u)E_{h}Y(u) \gamma(V) e^{\beta' Z}}{\big(E_{F}Y(u) \gamma(V) e^{\beta' Z}\big)^2}    \bigg\}.
\end{aligned}
\end{equation*}
where, $d_{F} \log P(T_i, \delta_i|\beta, \hat{\Lambda}_{\beta, F})$ represents the Hadamard derivative of $\log P(T_i, \delta_i|\beta, \hat{\Lambda}_{\beta, F})$ with respect to $F$ \citep{Hirose}.

%Here all derivatives are calculated treating $\gamma(V_i)$ as constant.\\
\textbf{Theorem 1:} At the true value of $(b, \beta, F)$, we are going to prove the followings

1. $\hat{\Lambda}_{\beta_0, F_0}(t)=\Lambda_0(t)$, the true cumulative hazard and

2. The score function $\phi(V, T, \delta| b_0, \beta_0, F_0)$ defined in (\ref{3.16})
is the efficient score function where we drop the subscript $i$.\\
\textbf{Proof:}
Replace $F_n$ by $F_0$,we get from (\ref{3.12})
\begin{equation}
\hat{\Lambda}_{\beta_0, F_0}(t)=\int_{0}^{t}\frac{E[ dN(u)]}{E[ \gamma(V) Y(u)e^{\beta_0' Z}]},
\end{equation}
where $E$ is the expectation with respect to the true distribution $F_0$. 
At the true value of the parameters $(\beta, F)$ we can write
\begin{equation} \label{3.20}
\begin{aligned}
E_{}[ dN(u)]={} & E_{}[ \gamma(V) Y(u)e^{\beta_0' Z}d\Lambda_0(u)]
%\\\hat{\Lambda}(t;\theta, F)={} & \int_{0}^{t} \frac{E_{} \gamma(V) Y(u)e^{\beta' Z}d\Lambda(u)}{E_{} \gamma(V) Y(u)e^{\beta' Z}}~~~~~~~~\mbox{(from equation 6.27)}\\
%\hat{\Lambda}(t;\theta, F)={} & \int_{0}^{t} d\Lambda(u)
%d\Lambda(u)={} &\frac{E_{} dN(u)}{E_{} \gamma(V) Y(u)e^{\beta' Z}}\\
%\Lambda(t)={} &\int_{0}^{t} \frac{E_{} dN(u)}{E_{} \gamma(V) Y(u)e^{\beta' Z}}
.\end{aligned}
\end{equation}

So from this point of view, we have $\hat{\Lambda}_{\beta_0, F_0}(t)=\Lambda_0(t).$

The score function $\phi(V, T, \delta|b, \beta, F)=\phi_l(V|b)+\phi_s(T, \delta|\beta, F)$ in (\ref{3.16}) has two parts. We know that the logistic model is a parametric model that does not involve $\Lambda$, so we will work on the survival part of the score function. So the score function for the survival part at the true value of the parameters $(\beta, F)$ can be expressed as
\begin{equation} \label{3.21}
\footnotesize
\begin{aligned}
\phi^*_\beta=\phi_s(T, \delta|\beta_0, F_0)={} &  \frac{\partial}{\partial \beta} \log P(T, \delta|\hat{\Lambda}_{\beta, F}, \beta) \bigg|_{\beta=\beta_0, F=F_0}
\\
%={} & \gamma(V) \frac{\partial}{\partial \beta} \bigg[\delta\big\{\log \frac{E dN(T)}{E \gamma(V) Y(T)e^{\beta' Z}}+\beta' Z\big\}-e^{\beta' Z}\int_{0}^{T}\frac{E dN(u)}{E \gamma(V) Y(u)e^{\beta' Z}}\bigg]\\
={} & \gamma(V) \bigg\{\delta \bigg[Z-\frac{E[ \gamma(V) Y(T)Ze^{\beta_0' Z}]}{E[\gamma(V) Y(T)e^{\beta_0' Z}]} \bigg]-e^{\beta_0' Z} \int_{0}^{T} \frac{E[ dN(u)]}{E[ \gamma(V) Y(u)e^{\beta_0' Z}]} \bigg[Z-\frac{E[ \gamma(V) Y(u)Ze^{\beta_0' Z}]}{E[ \gamma(V) Y(u)e^{\beta_0' Z}]} \bigg]     \bigg\}
.\end{aligned}
\end{equation}

Let $M_1(u)=E[ \gamma(V) Y(u)Ze^{\beta_0' Z}]$
and
$M_0(u)=E[ \gamma(V) Y(u)e^{\beta_0' Z}]$. So by using equation (\ref{3.20}), the above equation can be expressed as
\begin{equation} \label{3.22}
\begin{aligned}
\phi^*_\beta=\phi_s(T, \delta|\beta_0, F_0)
%={} & \gamma(V) \bigg\{\delta \bigg[Z-\frac{M_1(u)}{M_0(u)} \bigg]
%-e^{\beta' Z} \int_{0}^{T} \frac{E dN(u)}{M_0(u)} \bigg[Z-\frac{M_1(u)}{M_0(u)} \bigg]   \bigg\}\\
={} & \gamma(V) \bigg\{\delta \bigg[Z-\frac{M_1(T)}{M_0(T)} \bigg]
-e^{\beta_0' Z} \int_{0}^{T} \bigg[Z-\frac{M_1(u)}{M_0(u)} \bigg]   d \Lambda_0(u)\bigg\},
%={} & \gamma(V) \int_{0}^{\tau}  \bigg[Z-\frac{M_1(t)}{M_0(t)} \bigg] \bigg\{dN(t)- Y(t) e^{\beta' Z}d \Lambda(t)  \bigg\}\\
%={} & \gamma(V) \int_{0}^{\tau}  \bigg[Z-\frac{M_1(t)}{M_0(t)} \bigg] dM(t).
\end{aligned}
\end{equation}
which is the efficient score function for Cox PH cure model. The calculation of efficient score function based on the projection theory is given in Supplementary Materials (equation S5.12).\\
\textbf{3.3 Asymptotic Normality of the MLE}

\underline{Assumptions:}

To show the asymptotic normality of the MLE and its asymptotic variance, we have to consider some assumptions. On the set of cdf functions $\digamma$, we use the sup-norm, i.e., for $F, F_0 \in \digamma$,
$$||F-F_0 ||_\infty=\sup_x|F(x)-F_0(x)|. $$

For $\rho>0$, let
$$\zeta_\rho=\{F \in \digamma:||F-F_0 ||_\infty < \rho\} .$$
The assumptions are given below

$\textbf{A1:}$ We assume that there exists a finite number $\tau > 0$ such that $S(\tau)=P(T> \tau)=E[Y(\tau)]>0$.

$\textbf{A2:}$ The range of $Z$ is bounded and $\beta$ is in the compact set $\Theta$ which follows $||Z|| \leq M$ and $||\beta|| \leq M$ for some $0<M< \infty$.

$\textbf{A3:}$ The empirical cdf $F_n$ is $\sqrt{n}$ consistent i.e. $\sqrt{n} |F_n-F_0|=O_p(1)$.

$\textbf{A4:}$ The efficient information matrix $I_s^*=E[\phi^*_{\beta}\phi^{*'}_{\beta}]$ is invertible.\\
\textbf{Theorem-2}: If the assumptions (A1-A4) hold, then

1. $\hat{\beta}_n \stackrel{P}{\rightarrow} \beta_0$ as $n \rightarrow \infty$~~and~~
2. $\hat{\Lambda}_{\hat{\beta}_n, F_n}-\Lambda_0=o_p(1)$.

The proof of Theorem-2 is given in Supplementary Materials.\\
\textbf{Theorem-3:} The score functions $\phi_s(T, \delta|\beta, F)$ and $B(T, \delta|\beta, F)$ are defined previously. Suppose for $(A1)$-$(A4)$, $\hat{\beta}_n \stackrel{P}{\rightarrow} \beta_0$ and $F_n \stackrel{P}{\rightarrow} F_0$ as $n \rightarrow \infty$, then we have
\begin{equation*}
E\bigg[\sqrt{n}\bigg\{\phi_s(T, \delta|\hat{\beta}_n,F_0)-\phi_s(T, \delta|\beta_0,F_0) \bigg\}\bigg]=-E\bigg[\phi_s(T, \delta |\beta_0,F_0)\phi'_s(T, \delta |\beta_0,F_0)\bigg] \bigg\{\sqrt{n}(\hat{\beta}_n- \beta_0) \bigg\}+ o_p(1),
\end{equation*}
and 
\begin{equation*}
\begin{aligned}
E\bigg[\sqrt{n}\bigg\{\phi_s(T, \delta|\hat{\beta}_n,F_n)-\phi_s(T, \delta|\hat{\beta}_n,F_0) \bigg\}\bigg]
={} & -E\bigg[\phi_s(T, \delta|\beta_0,F_0)B(T, \delta|\beta_0,F_0)\bigg] \bigg\{\sqrt{n}(F_n- F_0) \bigg\}\\
& +o_p \big(1+\sqrt{n}(\hat{\beta}_n- \beta_0)\big).
\end{aligned}
\end{equation*}

\textbf{Remark}: The results are obtained without assuming the derivative of the score functions $\frac{\partial}{\partial \beta}\phi_s(T, \delta|\beta,F)$ and $d_F B(T, \delta|\beta,F)$ exist. This result give us asymptotic expansion of profile likelihood without differentiating the score function that involve implicit function.

The proof of Theorem-3 is given in Supplementary Materials.\\
\textbf{Theorem-4:} If the assumptions $\{A1, A2, A3, A4\}$ are satisfied, then a consistent estimator $\hat{\beta}_n$ to the estimating equation
\begin{equation*}
\sum_{i=1}^{n}\phi_s(T_i, \delta_i|\hat{\beta}_n, F_n)=0,
\end{equation*}
is an asymptotically linear estimator for $\beta_0$ \citep{Hirose} with the efficient influence function $(I_s^*)^{-1}\phi_s(T, \delta|\beta_0, F_0)$, so that
$$\sqrt{n} (\hat{\beta}_n-\beta_0)=\frac{1}{\sqrt{n}}\sum_{i=1}^{n}(I_s^*)^{-1}\phi_s(T_i, \delta_i|\beta_0, F_0)+o_p(1) \stackrel{D}{\longrightarrow}N\{0,(I_s^*)^{-1}\},$$
where $N\{0,(I_s^*)^{-1}\}$ is a normal distribution with mean zero and variance $(I_s^*)^{-1}$. So the estimator $\hat{\beta}_n$ is efficient.

The proof of Theorem-4 is given in Supplementary Materials.

\section{Simulation Study}
We are going to perform a simulation study where our goal is to compare and contrast the SMCURE package with our approach by assessing parameter estimation and standard error estimation. Survival times and censoring times were generated from Weibull proportional hazards model  and uniform distribution respectively. Simulation results for Cox PH cure model were evaluated with two covariates (fixed by design), one binary covariate from binomial distribution with probability 0.5 and one continuous covariate generated from standard normal distribution $N(0,1)$. Therefore, the covariate vectors for logistic and survival components were $W=(W_0,W_1,W_2)$ and $Z=(Z_1,Z_2)$ respectively.

The cure rates were varied through the coefficients ($b$) corresponding to $W$. The slight cure rate for the treatment group ($W_1=1$) was 25\% and for the control group ($W_1=0$) was 11\%, resulting from $b=(2.1,-1,0.3)$. The moderate cure rate for the treatment group was 50\% and for the control group was 27\%, resulting from $b=(1.022,-1,0.3)$. Moreover, The substantial cure rate for the treatment group was 75\% and for the control group was 53\%, resulting from $b=(-0.1,-1,0.3)$. For each configuration, mean was chosen (which is zero) as the value of the continuous covariate ($W_2$). Moreover, the coefficient vector for survival part was $\beta=(-1,0.5)$. These results includes a sample of 200 individuals ($n=200$) with 1000 replications from both SMCURE package and our approach.

\begin{table}
	\footnotesize
	\centering
	\caption{Simulation results for Cox PH cure model }
	\begin{tabular}{|l|l|l|l|l|l|l|l|l|l|l|l|l|}
		\hline
		\multicolumn{2}{|c|}{cure rate 25\%} &
		\multicolumn{4}{|c|}{SMCURE package} &
		\multicolumn{4}{c|}{Our approach} 
		\\ \hline
		Parameter & True value & Bias & SE & ESE &  CP & Bias & SE & ESE &  CP \\ \hline
		$b_0$ & 2.1 & 0.0622 & 0.3493 & 0.3570 & 0.9490 & 0.0678 & 0.3489 & 0.3429 & 0.9660 \\ \hline
		$b_1$ & -1 & -0.0270 & 0.4227 & 0.4255 & 0.9720 & -0.0527 & 0.4170 & 0.4182 & 0.9670 \\ \hline
		$b_2$ & 0.3 & 0.0104 & 0.2034 & 0.2116 & 0.9570 & 0.0414 & 0.1976 & 0.2025 & 0.9530 \\ \hline
		$\beta_1$ & -1 & 0.0019 & 0.1844 & 0.1834 & 0.9500 & 0.0152 & 0.1818 & 0.1707 & 0.9320 \\ \hline
		$\beta_2$ & 0.5 & 0.0055 & 0.0883 & 0.0924 & 0.9550 & -0.0073 & 0.0889 & 0.0908 & 0.9510 \\ \hline \hline
		\multicolumn{2}{|c|}{cure rate 50\%} &
		\multicolumn{4}{|c|}{SMCURE package} &
		\multicolumn{4}{c|}{Our approach} 
		\\ \hline
		Parameters & True value & Bias & SE & ESE &  CP & Bias & SE & ESE & CP \\ \hline
		$b_0$ & 1.022 & 0.0247 & 0.2352 & 0.2190 & 0.9290 & 0.0285 & 0.2353 & 0.2279 & 0.9480 \\ \hline
		$b_1$ & -1 & -0.0311 & 0.3084 & 0.3210 & 0.9590 & -0.0565 & 0.3082 & 0.3139 & 0.9550 \\ \hline
		$b_2$ & 0.3 & 0.0091 & 0.1691 & 0.1763 & 0.9580 & 0.0211 & 0.1697 & 0.1712 & 0.9570 \\ \hline
		$\beta_1$ & -1 & -0.0152 & 0.2240 & 0.2232 & 0.9530 & -0.0086 & 0.2241 & 0.2035 & 0.9300 \\ \hline
		$\beta_2$ & 0.5 & 0.0103 & 0.1155 & 0.1153 & 0.9490 & -0.0003 & 0.1147 & 0.1172 & 0.9530 \\ \hline \hline
		\multicolumn{2}{|c|}{cure rate 75\%} &
		\multicolumn{4}{|c|}{SMCURE package} &
		\multicolumn{4}{c|}{Our approach} 
		\\ \hline
		Parameters & True value & Bias & SE & ESE &  CP & Bias & SE & ESE &  CP \\ \hline
		$b_0$ & -0.1 & -0.0049 & 0.2043 & 0.1926 & 0.9370 & 0.0105 & 0.2109 & 0.2073 & 0.9430 \\ \hline
		$b_1$ & -1 & -0.0106 & 0.3200 & 0.3283 & 0.9640 & -0.0315 & 0.3362 & 0.3231 & 0.9490 \\ \hline
		$b_2$ & 0.3 & 0.0027 & 0.1603 & 0.1603 & 0.9530 & 0.0191 & 0.1678 & 0.1580 & 0.9440 \\ \hline
		$\beta_1$ & -1 & -0.0092 & 0.3112 & 0.3240 & 0.9580 & 0.0186 & 0.3207 & 0.2948 & 0.9150 \\ \hline
		$\beta_2$ & 0.5 & 0.0147 & 0.1401 & 0.1449 & 0.9530 & -0.0045 & 0.1427 & 0.1484 & 0.9600 \\ \hline
	\end{tabular}
\end{table}
The results from simulation studies such as estimate biases, standard errors (SE), estimated standard errors (ESE) and confidence interval coverage probabilities (CP) for each configuration are given in Table-1. The standard errors (SE) have been calculated using 1000 simulation estimates. In SMCURE packgae, bootstrap samples have been used to compute ESE whereas in our approach we have calculated the ESE analytically through the profile likelihood score function.
For coverage probabilities (CP), we have computed the 95\% confidence interval for each of the parameter estimates and determine the frequency in which the true parameter value was captured.

From Table 1, we can observe that for both SMCURE package and our approach, the parameter estimates are close to the true values and estimate biases are very small with most less than 0.05. For all configurations, with only a few exceptions, the SE and ESE of the parameters are very close for both SMCURE package and our approach. The capture rates based on the confidence interval are relatively similar for SMCURE package and our approach. 

Due to the complexity of the estimating equation in SMCURE package, the ESE of estimated parameters are not directly available. As a result, the package used bootstrap samples to compute the standard errors of estimated parameters. On the other hand, we have found the explicit form of the efficient score function (via the profile likelihood score function) and hence computed the ESE analytically through the efficient information matrix.

\begin{comment}

\begin{table}
\footnotesize
\centering
\caption{Simulation results for n=200 and number of simulation=100 (cure rate= 25\%)}
\begin{tabular}{|l|l|l|l|l|l|l|l|l|l|l|l|l|}
\hline
\multicolumn{2}{|c|}{} &
\multicolumn{4}{|c|}{SMCURE package} &
\multicolumn{4}{c|}{Our approach} 
\\ \hline
Parameters & True value & Bias & SE & ESE & 95\% CP & Bias & SE & ESE & 95\% CP \\ \hline
b0 & 1.022 & 0.0247 & 0.2352 & 0.2190 & 0.9290 & 0.0285 & 0.2353 & 0.2279 & 0.9480 \\ \hline
b1 & -1 & -0.0311 & 0.3084 & 0.3210 & 0.9590 & -0.0565 & 0.3082 & 0.3139 & 0.9550 \\ \hline
b2 & 0.3 & 0.0091 & 0.1691 & 0.1763 & 0.9580 & 0.0211 & 0.1697 & 0.1712 & 0.9570 \\ \hline
beta-1 & -1 & -0.0152 & 0.2240 & 0.2232 & 0.9530 & -0.0086 & 0.2241 & 0.2035 & 0.9300 \\ \hline
beta-2 & 0.5 & 0.0103 & 0.1155 & 0.1153 & 0.9490 & -0.0003 & 0.1147 & 0.1172 & 0.9530 \\ \hline

\end{tabular}
\end{table}

\end{comment}

\section{Application to Eastern Cooperative Oncology Group (ECOG) Data}
We have used the melanoma data (ECOG phase III clinical trial e1684) from SMCURE package \citep{Cai et al.} as a numerical example to compare our results with the output obtained from SMCURE package. The advantage of our approach is that we have used the efficient score function to get the standard errors of the estimated parameters whereas in SMCURE package, bootstrap sampling procedure has been used due to the complexity of the estimating equation in the EM algorithm (implicit form of their score function).

In the dataset, the subjects had melanoma cancer and were treated with interferon alpha-2b (IFN) regimen. The purpose of the study was to investigate the effects of  high dose interferon alpha-2b (IFN) regimen against the placebo as the postoperative adjuvant therapy. In this example, relapse free survival is defined as the event and the time from initial treatment to recurrence of melanoma is defined as failure time. A total number of 284 observations (after deleting two missing observations) has been used for the statistical analysis. Three covariates are considered:  gender (0=male,1=female), treatment (0=control,1=treatment) and age (continuous variable which is centered to the mean) for both the incidence and latency parts.

Out of 284 individuals, 196 had melanoma cancer recurring (approximately 31\% censoring rate). The observed follow-up time of the individuals ranged from 0.032 to 9.643 years. 
%Among the events, there were 137 distinct event times and 25 event times with ties (where ties ranged from 2 to 4). 
The parameter estimates, standard errors and 95\% CI using SMCURE package and our approach (for logistic and Cox PH components) are given in Table 2 and Table 3 respectively.

From Table 2, we observed that in SMCURE package only intercept was significant at 5\% level of significance whereas in our approach, intercept and treatment both have significant effects in determining the long term incidence. The result for treatment suggests that the probability of recurring melanoma for control group is significantly higher compared to the treatment group.
%because the recurrence of melanoma is determined by the growth rate of the surviving cancer cells which is potentially identified by patient specific factors such as age.
However, age and sex both are insignificant on SMCURE package and our approach.
\begin{table}[t!] %***
	\footnotesize
	\caption{Results for logistic component from ECOG Data}
	\label{tab:simulation}\par
	\vskip .1cm
	\centerline{\tabcolsep=3truept
		\begin{tabular}{|l|lll|lll|}
			\hline
			\multirow{2}{*}{Covariates} &
			\multicolumn{3}{c|}{SMCURE package} &
			\multicolumn{3}{c|}{Our approach}
			\\
			& Estimates &~SE & ~~~~95\% CI & Estimates &~SE & ~~~~95\% CI \\
			\hline
			Intercept&1.3649&0.2877&(0.8012,1.9286)& 1.3703&0.2698&(0.8414, 1.8991)\\
			Treatment&-0.5884&0.3065&(-1.1889,0.0121)&-0.6712&0.3026&(-1.2643, -0.0780)\\
			Age&0.0203&0.0145&(-0.0081,0.0487)&0.0159&0.0115&(-0.0066, 0.03844)\\
			Sex&-0.0869&0.3291&(-0.7319,0.5581)&-0.0361&0.3026&(-0.6291, 0.5569)\\
			\hline
	\end{tabular}}
\end{table}

On the other hand, from Table 3 it is observed that in both SMCURE package and our approach, all the covariates have insignificant effect on latency.

\begin{table}[t!] %***
	\footnotesize
	\caption{Results for Cox PH component from ECOG Data}
	\label{tab:simulation}\par
	\vskip .1cm
	\centerline{\tabcolsep=3truept
		\begin{tabular}{|l|lll|lll|}
			\hline
			\multirow{2}{*}{Covariates} &
			\multicolumn{3}{c|}{SMCURE package} &
			\multicolumn{3}{c|}{Our approach} 
			\\
			& Estimates &~~SE & ~~~~95\% CI & Estimates &~~SE &~~~~ 95\% CI \\
			\hline
			Treatment&-0.1535&0.1721&(-0.4908,0.1838)&-0.0999&0.1540&(-0.4017, 0.2019)  \\
			Age&-0.0077&0.0067&(-0.0208,0.0005)&-0.0044&0.0069&(-0.0181, 0.0093)  \\
			Sex&0.0995&0.1908&(-0.2744,0.4734)&0.0460&0.1542&(-0.2562, 0.3482)  \\
			\hline
	\end{tabular}}
\end{table}

%\pagebreak
%\begin{figure}[h!]
%	\centering
%	\includegraphics[width=5in]{Rplot.pdf}
%\end{figure}
%A positive non-significant estimate value for gender on latency indicates earlier recurrence times of melanoma for females compared to males. Similarly, insignificant estimate for treatment groups indicates later recurrence times for treatment compared to placebo.

\section{Discussion}
Over the years, many techniques have been used which can avoid differentiation of the implicit function under the profile likelihood function (\citealp{Murphy and Vaart,Zeng and Cai,Zeng and Lin,Lu,Zeng and Lin (a)}). Therefore, these approaches didn't involve the score function for the profile likelihood. As a result, the asymptotic variance of the estimator has shown without using the profile likelihood score function.
%they didn't derive the profile likelihood score function to  show the asymptotic variance of the estimator as the  inverse of the efficient information matrix. Our contribution in this paper is that we have expressed the efficient information matrix as the variance of the profile likelihood score function.

In this paper, we have shown the asymptotic normality of the maximum profile likelihood estimator via asymptotic expansion of the profile likelihood and compute the efficient information matrix based on the profile likelihood score function. This is an additional method to compute standard errors for the maximum profile likelihood estimator.

%obtain the explicit form of the variance estimator with an implicit function in the profile likelihood. we have also shown the profile likelihood score function and the efficient score function based on projection theory are equal. Moreover, we have used the profile likelihood score function to compute efficient information matrix. Therefore, our approach provides an additional method to compute standard errors from the profile likelihood score function.

%For simulation study and real-life application (melanoma data from ECOG phase III clinical trial e1684), we have used the profile likelihood score function to compute the standard errors of the estimated parameters which provides competitive results to SMCURE package (using bootstrap samples for variance estimation).

%%%%%%%%%%%%%%%%%%%%%%%%%%%%%%%%%%%%%%%%%%%%%%%%%%%%%%%%%%%%%%%%%%%%%%%%%%%%%%%%%%%%%%%%%%%%%%%%%%%%%%%%%%%%%%%%%%%%%%%%%%%%
%\vskip 14pt
\section*{Supplementary Materials}
An additional document has been provided as Supplementary Materials, where proofs of all necessary Theorems have given.

\par
%%%%%%%%%%%%%%%%%%%%%%%%%%%%%%%%%%%%%%%%%%%%%%%%%%%%%%%%%%%%%%%%%%%%%%%%%%%%%%%%%%%%%%%%%%%%%%%%%%%%%%%%%%%%%%%%%%%%%%%%%%%%
%\vskip 14pt
%\noindent {\large\bf Acknowledgements}

%Write the acknowledgements here.
\par

%%%%%%%%%%%%%%%%%%%%%%%%%%%%%%%%%%%%%%%%%%%%%%%%%%%%%%%%%%%%%%%%%%%%%%%%%%%%%%%%%%%%%%%%%%%%%%%%%%%%%%%%%%%%%%%%%%%%%%%%%%%%
\markboth{\hfill{\footnotesize\rm KHANDOKER AKIB MOHAMMAD AND OTHERS} \hfill}
{\hfill {\footnotesize\rm EFFICIENT ESTIMATION FOR COX PH CURE MODEL} \hfill}

%\iffalse
\bibhang=1.7pc
\bibsep=2pt
\fontsize{9}{14pt plus.8pt minus .6pt}\selectfont
\renewcommand\bibname{\large \bf References}
%\begin{thebibliography}{11}
\expandafter\ifx\csname
natexlab\endcsname\relax\def\natexlab#1{#1}\fi
\expandafter\ifx\csname url\endcsname\relax
\def\url#1{\texttt{#1}}\fi
\expandafter\ifx\csname urlprefix\endcsname\relax\def\urlprefix{URL}\fi
%\fi

\lhead[\footnotesize\thepage\fancyplain{}\leftmark]{}\rhead[]{\fancyplain{}\rightmark\footnotesize{} }%Put this line in Page 2
%%%%%%%%%%%%%%%%%%%%%%%%%%%%%%%%%%%%%%%%%%%%%%%%%%%%%%%%%%

\pagebreak
\renewcommand{\baselinestretch}{2}

\markright{ \hbox{\footnotesize\rm  Supplement
		%{\footnotesize\bf 24} (201?), 000-000
	}\hfill\\[-13pt]
	\hbox{\footnotesize\rm
		%\href{http://dx.doi.org/10.5705/ss.20??.???}{doi:http://dx.doi.org/10.5705/ss.20??.???}
	}\hfill }

\markboth{\hfill{\footnotesize\rm KHANDOKER AKIB MOHAMMAD AND OTHERS} \hfill}
{\hfill {\footnotesize\rm EFFICIENT ESTIMATION FOR COX PH CURE MODEL} \hfill}

\renewcommand{\thefootnote}{}
$\ $\par \fontsize{10}{10pt plus.8pt minus .6pt}\selectfont

%%%%%%%%%%%%%%%%%%%%%%%%%%%%%%%%%%%%%%%%%%%%%%%%%%%%%%%%%%%%%%%%%%%%%%%%%%%%%%%%%%%%%%%%%%%%%%%%%%%%%%%%%%%%%%%%%%%%%%%%%%%%

\centerline{\large\bf EFFICIENT ESTIMATION FOR THE}
\vspace{2pt}
\centerline{\large\bf COX PROPORTIONAL HAZARDS CURE MODEL}
\vspace{.25cm}
%\author{Author(s)}
%\vspace{.4cm}
\centerline{ Khandoker Akib Mohammad$^\ast$, Yuichi Hirose, Budhi Surya and Yuan Yao}
\vspace{.25cm}
\centerline{\it Victoria University of Wellington}
\vspace{.55cm}
\centerline{\bf Supplementary Material}
\vspace{.55cm}
\fontsize{9}{11pt plus.8pt minus .6pt}\selectfont
\noindent

%CONTENT OF A BRIEF NOTE.........\\
%CONTENT OF THE BRIEF NOTE.\\
%CONTENT OF THE BRIEF NOTE.\\
\par

\setcounter{section}{0}
\setcounter{equation}{0}
\def\theequation{S\arabic{section}.\arabic{equation}}
\def\thesection{S\arabic{section}}

\fontsize{10}{10pt plus.8pt minus .6pt}\selectfont
\normalsize
\section{Lemma with Proof}
\textbf{Lemma-1}: Let $\digamma$ be the set of cdf functions and $\zeta_\rho \subset \digamma$ ($\zeta_\rho$ is defined in Section-3.3 of the main manuscript). If the assumptions (\textbf{A1-A4}) hold, then 

(i) $P(T, \delta|\beta, F)$ is bounded away from zero.

(ii) The class of functions $\bigg\{ \log P \big(T, \delta|\beta, F \big): \beta \in \Theta, F \in \zeta_\rho \bigg\} \mbox{ is uniformly bounded Donsker.}$

(iii) The class of functions $\bigg\{ \phi_s \big(T, \delta|\beta, F \big): \beta \in \Theta, F \in \zeta_\rho \bigg\} \mbox{ is uniformly bounded Donsker.}$\\
\textbf{Proof}: For (i), we know

\begin{equation} \label{S1.1}
P(T, \delta|\beta, F)= \bigg[\frac{E_{F} {d N(T)}}{E_{F} \gamma(V) Y(T)e^{\beta' Z}} e^{\beta' Z}\bigg]^{\delta \gamma(V)}  \bigg\{  \bigg[\exp{- \int_{0}^{T}\frac{E_{F} dN(u)}{E_{F} \gamma(V) Y(u)e^{\beta' Z}}}\bigg]^{e^{\beta' Z}}  \bigg\}^ {\gamma(V)}.
\end{equation}

Since the map $(f, F) \rightarrow E_F(f)=\int f dF$ is continuous, there is a constant $c>0$, such that for all $F \in \zeta_\rho$ (based on $\textbf{A1}$), we can write
$$E_F[Y(\tau)] \geq c>0 .$$

We know $\gamma(V)=\bigg(\frac{p S(\tau)}{1-p+p S(\tau)} \bigg)^{1-\delta}$, so we have
\begin{equation} \label{S1.2}
0< \frac{c p}{1-p+c p} \leq \gamma(V) \leq 1.
\end{equation} 

On the basis of $\textbf{A2}$, we can write $e^{-M^2} \leq e^{\beta'Z} \leq e^{M^2}$. So the upper bound of $E_{F} \gamma(V) Y(u)e^{\beta' Z}$ can be expressed as
\begin{equation}
\big | E_{F} \gamma(V) Y(u)e^{\beta' Z} \big | \leq \big | E_{F} e^{\beta' Z} \big | \leq e^{M^2}.
\end{equation} 

Now by using equation (\ref{S1.2}), we can write
\begin{equation} \label{s1.4}
E_{F} \gamma(V) Y(u)e^{\beta' Z} \geq  \frac{c^2 p}{1-p+c p}e^{-M^2} >0.
\end{equation}

For some constant $c_1 >0$, we can write $0< c_1 \leq E_{F}{d N(u)} \leq 1$. Since $E_{F} \gamma(V) Y(u)e^{\beta' Z}$ is bounded away from zero (equation \ref{s1.4}), we get
$$ 0< \frac{c_1}{e^{M^2}} \leq \frac{E_{F} {d N(u)}}{E_{F} \gamma(V) Y(u)e^{\beta' Z}} \leq \frac{ e^{M^2} (1-p+ c p)  }{c^2 p }. $$

When $\delta=1$, from (\ref{S1.1}) we get
\begin{equation*}
\begin{aligned}
P(T, \delta|\beta, F)={} &  \bigg[\frac{E_{F} {d N(T)}}{E_{F} \gamma(V) Y(T)e^{\beta' Z}} e^{\beta' Z}\bigg] \bigg[\exp{- \int_{0}^{T}\frac{E_{F} dN(u)}{E_{F} \gamma(V) Y(u)e^{\beta' Z}}}\bigg]^{e^{\beta' Z}}\\
P(T, \delta|\beta, F)  \geq {} &  \frac{c_1}{e^{M^4}}  \bigg\{ e^{ - \frac{c_1}{e^{M^4}}} \bigg\}, 
\end{aligned}
\end{equation*}
and when $\delta=0$,
\begin{equation*}
\begin{aligned}
P(T, \delta|\beta, F)={} &  \bigg\{  \bigg[\exp{- \int_{0}^{T}\frac{E_{F} dN(u)}{E_{F} \gamma(V) Y(u)e^{\beta' Z}}}\bigg]^{e^{\beta' Z}}  \bigg\}^ {\gamma(V)}\\
P(T, \delta|\beta, F)  \geq {} & \bigg\{ e^{ - \frac{c_1 c p}{e^{M^4}(1-p+ c p)}} \bigg\}.
\end{aligned}
\end{equation*}

From the above equations, we can write
\begin{equation}
P(T, \delta|\beta, F) \geq  \bigg\{ e^{ - \frac{c_1 c p}{e^{M^4}(1-p+ c p)}} \bigg\} > 0.
\end{equation}

So finally we can say that $P(T, \delta|\beta, F)$ is bounded away from zero and hence (i) is proved.\\	
For (ii), the profile log-likelihood function of the survival part for Cox PH cure model is
$$ \log P(T, \delta|\beta,F)= \gamma(V)\bigg[\delta \big\{\log \frac{E_{F} dN(T)}{E_{F} \gamma(V) Y(T)e^{\beta' Z}}+\beta' Z \big\}-e^{\beta'Z} \int_{0}^{T}\frac{E_{F} dN(u)}{E_{F} \gamma(V) Y(u)e^{\beta' Z}} \bigg]. $$

We know the set of cdf functions $\digamma$ is uniformly bounded Donsker. Hence the subset $\zeta_\rho \subset \digamma$ is uniformly bounded Donsker. The class of functions $\big\{N(t): t \in [0, \tau] \big\}$ and $\big\{Y(t): t \in [0, \tau] \big\}$ are uniformly bounded Donsker (Theorem 2.10.6 in Van Der Vaart and Wellner, 1996). 

The class of functions $\big\{\beta'Z: \beta \in \Theta \big\}$ is Lipschitz in $\beta$. So, by Theorem 2.10.6 in Van Der Vaart and Wellner (1996), the class of functions $\big\{\beta'Z: \beta \in \Theta \big\}$ is uniformly bounded Donsker. 

%Since $||Z|| \leq M$ and $||\beta|| \leq M$, so we can write ${-M^2} \leq {\beta' Z} \leq {M^2}$.  For the class of functions $\big\{\beta'Z: \beta \in \Theta \big\}$ with compact set $\Theta$, we can write
%$$ \bigg|\beta'_1 Z-\beta'_2 Z \bigg| \leq M \bigg(\big|\big|\beta_1-\beta_2 \big|\big|\bigg),$$
%which is Lipschitz in $\beta$. So, by Theorem 2.10.6 in Van der Vaart and Wellner, the class of functions $\big\{\beta'Z: \beta \in \Theta \big\}$ is uniformly bounded Donsker. 

%Since $e^{Z}$ is a Lipschitz continuous function, we have
%$$ \bigg|e^{\beta'_1 Z}-e^{\beta'_2 Z} \bigg|  \leq M e^{M^2} \bigg(\big|\big|\beta_1-\beta_2 \big|\big|\bigg),$$
%so by Theorem 2.10.6 in Van der Vaart and Wellner, the class of functions $\big\{e^{\beta' Z}: \beta \in \Theta \big\}$ is uniformly bounded Donsker.

Since $e^{\beta' Z}$ is a Lipschitz continuous function, so by Theorem 2.10.6 in Van Der Vaart and Wellner (1996), the class of functions $\big\{e^{\beta' Z}: \beta \in \Theta \big\}$ is uniformly bounded Donsker.

Since $\big\{Y(t): t \in [0, \tau] \big\}$ and $\big\{e^{\beta' Z}: \beta \in \Theta \big\}$ are uniformly bounded Donsker, so by Example 2.10.8 (Van Der Vaart and Wellner, 1996), the class of functions $\big\{\gamma(V)Y(t) e^{\beta' Z}: t \in [0, \tau], \beta \in \Theta \big\}$ is uniformly bounded Donsker. 

Since $ E_F(f)=\int f dF$ is Lipschitz, so for the class of functions $\big\{E_{F} \gamma(V) Y(t)e^{\beta' Z}: t \in [0, \tau], \beta \in \Theta, F \in \zeta_\rho \big\}$, we can write
\begin{equation*}
\normalsize
\begin{aligned}
\bigg |E_{F_1} \gamma(V) Y(t)e^{\beta'_1 Z}- E_{F_2} \gamma(V) Y(t)e^{\beta'_2 Z} \bigg | ={} & \int \gamma(V) Y(t) \bigg |e^{\beta'_1 Z}-e^{\beta'_2 Z} \bigg | dF_1 + \int \gamma(V) Y(t)e^{\beta_2' Z} d \big|\big|F_1-F_2  \big|\big|\\
\leq & \int \bigg |e^{\beta'_1 Z}-e^{\beta'_2 Z} \bigg | dF_1 + \int e^{\beta_2' Z} d \big|\big|F_1-F_2  \big|\big|\\
%\leq & \int || Ze^{\beta'_ Z}||  dF \big|\big|\beta_1-\beta_2  \big|\big| + \int e^{\beta' Z} d \big|\big|F_1-F_2  \big|\big|\\
\leq & ~M e^{M^2} \big|\big|\beta_1-\beta_2  \big|\big| + e^{M^2} \big|\big|F_1-F_2  \big|\big|.
\end{aligned}
\end{equation*}

Let max$ \bigg(M e^{M^2}, e^{M^2} \bigg)=M e^{M^2}$, then the above equation can be expressed as
\begin{equation}
\bigg |E_{F_1} \gamma(V) Y(t)e^{\beta'_1 Z}- E_{F_2} \gamma(V) Y(t)e^{\beta'_2 Z} \bigg | \leq M e^{M^2} \bigg( \big|\big|\beta_1-\beta_2  \big|\big| + \big|\big|F_1-F_2  \big|\big|\bigg),
\end{equation}
which is Lipschitz in parameters ($\beta$, F). So by Theorem 2.10.6 in Van Der Vaart and Wellner (1996), the class of functions $\big\{E_{F} \gamma(V) Y(t)e^{\beta' Z}: t \in [0, \tau], \beta \in \Theta, F \in \zeta_\rho \big\}$ is uniformly bounded Donsker. Similarly the class of functions $\big\{E_{F}N(t): t \in [0, \tau] \big\}$ is uniformly bounded Donsker.

Since $\big\{E_{F} \gamma(V) Y(t)e^{\beta' Z}: t \in [0, \tau], \beta \in \Theta, F \in \zeta_\rho \big\}$ is uniformly bounded Donsker and $E_{F} \gamma(V) Y(u)e^{\beta' Z}$ is bounded away from zero (equation \ref{s1.4}), by Example 2.10.9 in Van Der Vaart and Wellner (1996), the class of functions $$\bigg\{\frac{1}{E_F \gamma(V) Y(t) e^{\beta'Z}}: t \in [0, \tau], \beta \in \Theta, F \in \zeta_\rho \bigg\}$$
is uniformly bounded Donsker.

Since the map $(f, F) \rightarrow E_F(f)=\int f dF$ is Lipschitz, by Theorem 2.10.6 (Van Der Vaart and Wellner, 1996), the class of functions
$$\bigg\{\int_{0}^{t} \frac{E_FdN(u)}{E_F \gamma(V) Y(u) e^{\beta'Z}}: t \in [0, \tau], \beta \in \Theta, F \in \zeta_\rho \bigg\}$$
is uniformly bounded Donsker.

Since $\big\{e^{\beta' Z}: \beta \in \Theta \big\}$ is uniformly bounded Donsker, so by Example 2.10.8 (Van Der Vaart and Wellner, 1996), the class of functions $$\bigg\{ e^{\beta'Z} \int_{0}^{t} \frac{E_{F} dN(u)}{E_{F} \gamma(V) Y(u)e^{\beta' Z}} : t \in [0, \tau], \beta \in \Theta, F \in \zeta_\rho \bigg\}$$
is uniformly bounded Donsker. 

Since the class $\big\{{\beta' Z}: \beta \in \Theta \big\}$ is uniformly bounded Donsker, by Example 2.10.7 (Van Der Vaart and Wellner, 1996), the class of functions
$$\bigg\{ \log \frac{E_FdN(T)}{E_F \gamma(V) Y(T) e^{\beta'Z}}+ {\beta'Z}: \beta \in \Theta, F \in \zeta_\rho \bigg\}$$
is uniformly bounded Donsker.

Since the map $(f, F) \rightarrow E_F(f)=\int f dF$ is Lipschitz, so by Theorem 2.10.6 in Van Der Vaart and Wellner (1996), the class of functions $\bigg\{ \log P(T, \delta|\beta,F): \beta \in \Theta, F \in \zeta_\rho \bigg\}$ is uniformly bounded Donsker. So (ii) is proven.

For (iii), we know the score function of the survival part for Cox PH cure model is
\begin{equation*}
\small
\phi_s(T, \delta|\beta, F)= \gamma(V) \bigg\{\delta \bigg[Z-\frac{E_{F} \gamma(V) Y(T)Ze^{\beta' Z}}{E_{F} \gamma(V) Y(T)e^{\beta' Z}} \bigg]-e^{\beta' Z} \int_{0}^{T} \frac{E_{F} dN(u)}{E_{F} \gamma(V) Y(u)e^{\beta' Z}} \bigg[Z-\frac{E_{F} \gamma(V) Y(u)Ze^{\beta' Z}}{E_{F} \gamma(V) Y(u)e^{\beta' Z}} \bigg]    \bigg\}.
\end{equation*}

Similar proof to (ii), we can show that the class of functions $\bigg\{ \phi_s(T, \delta|\beta, F): \beta \in \Theta, F \in \zeta_\rho \bigg\}$ is uniformly bounded Donsker.\\
\textbf{Lemma-2}: If the assumptions (\textbf{A1-A4}) hold, then 
\begin{equation*}
\begin{aligned}
\bigg|\bigg| \phi_s(T, \delta|\hat{\beta}_n,F_n)- \phi_s(T, \delta|\beta_0,F_0) \bigg|\bigg|%={} & \bigg|\bigg| S_s(T, \delta|\hat{\beta}_n,F_n, \hat{\Lambda}_{\hat{\beta}_n,F_n} )- S_s(T, \delta|\beta_0,F_0, \hat{\Lambda}_{{\beta}_0,F_0}) \bigg|\bigg| \\
\leq &~ M'' \bigg( \big|\big| \hat{\beta}_n- \beta_0   \big|\big|+\big|\big|F_n-F_0 \big|\big|+ \big|\big| \hat{\Lambda}_{\hat{\beta}_n, F_n}- \hat{\Lambda}_{{\beta}_0, F_0} \big|\big|   \bigg),
\end{aligned}
\end{equation*}
where $M''$ is a $P_0$-square integrable function.\\
\textbf{Proof}: From equation (3.18) of the main manuscript, the score function for the survival part is
\begin{equation*}
\begin{aligned}
\phi_s(T, \delta|\beta, F_n)%={} & S_s \big(T, \delta|\beta, F_n,\hat{\Lambda}_{{\beta}, F_n} \big)\\
={} & \gamma(V) \bigg\{\delta \bigg[Z-\frac{E_{F_n} \gamma(V) Y(T)Ze^{\beta' Z}}{E_{F_n} \gamma(V) Y(T)e^{\beta' Z}} \bigg]-e^{\beta' Z} \int_{0}^{T}  \bigg[Z-\frac{E_{F_n} \gamma(V) Y(u)Ze^{\beta' Z}}{E_{F_n} \gamma(V) Y(u)e^{\beta' Z}} \bigg] d\hat{\Lambda}_{\beta, F_n}(u)    \bigg\}.
\end{aligned}
\end{equation*}
Define
\begin{equation*}
\begin{aligned}
\phi_s(T, \delta|\beta, F, \Lambda)
%={} & S_s \big(T, \delta|\beta, F_n,\hat{\Lambda}_{{\beta}, F_n} \big)\\
={} & \gamma(V) \bigg\{\delta \bigg[Z-\frac{E_{F} \gamma(V) Y(T)Ze^{\beta' Z}}{E_{F} \gamma(V) Y(T)e^{\beta' Z}} \bigg]-e^{\beta' Z} \int_{0}^{T}  \bigg[Z-\frac{E_{F} \gamma(V) Y(u)Ze^{\beta' Z}}{E_{F} \gamma(V) Y(u)e^{\beta' Z}} \bigg] d{\Lambda}(u)    \bigg\}.
\end{aligned}
\end{equation*}
Then the function is differentiable with respect to $\beta$, $F$ and $\Lambda$. Now we have
$$\phi_s(T, \delta|\beta, F_n)=\phi_s \big(T, \delta|\beta, F_n,\hat{\Lambda}_{{\beta}, F_n} \big). $$

Similar to the proof of Lemma-1, we can show that the derivative of the score function will also be uniformly bounded. 

From these we can say that the class of functions $\big\{ \phi_s(T, \delta|\beta, F, \Lambda): \beta \in \Theta, F \in \zeta_\rho, \Lambda \in H \big\}$ is Lipschitz in parameters $(\beta, F, \Lambda)$ and the result follows: 
\begin{equation*}
\begin{aligned}
\bigg|\bigg| \phi_s(T, \delta|\hat{\beta}_n,F_n)- \phi_s(T, \delta|\beta_0,F_0) \bigg|\bigg|={} & \bigg|\bigg| \phi_s(T, \delta|\hat{\beta}_n,F_n, \hat{\Lambda}_{\hat{\beta}_n,F_n} )- \phi_s(T, \delta|\beta_0,F_0, \hat{\Lambda}_{{\beta}_0,F_0}) \bigg|\bigg| \\
\leq &~ M'' \bigg( \big|\big| \hat{\beta}_n- \beta_0   \big|\big|+\big|\big|F_n-F_0 \big|\big|+ \big|\big| \hat{\Lambda}_{\hat{\beta}_n, F_n}- \hat{\Lambda}_{{\beta}_0, F_0} \big|\big|   \bigg).
\end{aligned}
\end{equation*}
%where $M''$ is a $P_0$-square integrable function.

\section{Theorem- 2 with proof}
\setcounter{equation}{0}

\textbf{Theorem 2}: If the assumptions (\textbf{A1-A4}) hold, then

1. $\hat{\beta}_n \stackrel{P}{\rightarrow} \beta_0$ as $n \rightarrow \infty$~~and~~

2. $\hat{\Lambda}_{\hat{\beta}_n, F_n}-\Lambda_0=o_p(1)$\\
\textbf{Proof}: For (1), we are going to use the idea of Theorem-5.7 (Van der Vaart, 2000), where we have to show

(i) $ \int \log P(T, \delta|\hat{\beta}_n, F_n)dF_n -\int \log P(T, \delta|\hat{\beta}_n, F_0) dF_0 \stackrel{P}{\longrightarrow} 0$~~ as $n \rightarrow \infty$

(ii) $E \bigg[\log \frac{P(T, \delta| \beta_0, F_0)}{P(T, \delta|\hat{\beta}_n, F_0)} \bigg]> 0$~~ if $\beta_0 \neq \hat{\beta}_n$

We will start with (i). Since the class of functions $\bigg\{ \log P(T, \delta|\beta, F): \beta \in \Theta, F \in \zeta_\rho \bigg\}$
is uniformly bounded Donsker. Hence it is Glivenko-Cantelli. So we can write
\begin{equation*}
\int \log P(T, \delta|\hat{\beta}_n, F_n)dF_n -\int \log P(T, \delta|\hat{\beta}_n, F_0) dF_0 \stackrel{P}{\longrightarrow} 0~~ \mbox{as}~~ n \rightarrow \infty
\end{equation*}

For (ii), we are going to use the idea of Kullback-Leibler (KL) distance. The distance between $P(T, \delta| \beta_0, F_0)$ and $P(T, \delta|\hat{\beta}_n, F_0)$ can be written as
\begin{equation} \label{s2.1}
E \bigg[\log \frac{P(T, \delta| \beta_0, F_0)}{P(T, \delta|\hat{\beta}_n, F_0)} \bigg]= \int dP(T, \delta| \beta_0, F_0)~\log \frac{P(T, \delta| \beta_0, F_0)}{P(T, \delta|\hat{\beta}_n, F_0)}.
\end{equation}

We know that $-\log x$ is a convex function for $x > 0$, so by using Jensen's inequality in (\ref{s2.1}), we can write
\begin{equation}
\begin{aligned}
E \bigg[\log \frac{P(T, \delta| \beta_0, F_0)}{P(T, \delta| \hat{\beta}_n, F_0)} \bigg]={} & \int dP(T, \delta| \beta_0, F_0)~\log \frac{P(T, \delta| \beta_0, F_0)}{P(T, \delta|\hat{\beta}_n, F_0)}\\
={} & - \int dP(T, \delta| \beta_0, F_0)~\log \frac{P(T, \delta|\hat{\beta}_n, F_0)}{P(T, \delta| \beta_0, F_0)}\\
> & - \log \int  dP(T, \delta| \beta_0, F_0)~\frac{P(T, \delta|\hat{\beta}_n, F_0)}{P(T, \delta| \beta_0, F_0)}\\
={} & - \log \int dP(T, \delta|\hat{\beta}_n, F_0)\\
={} & - \log 1\\
={} & 0~~~~~~~~~~~;\mbox{if}~~ \hat{\beta}_n \neq \beta_0
\end{aligned}
\end{equation}
and
$$E \bigg[\log \frac{P(T, \delta| \beta_0, F_0)}{P(T, \delta|\hat{\beta}_n, F_0)} \bigg]=0~~\mbox{if}~~\beta_0  = \hat{\beta}_n .$$

Hence (ii) is also proven.

So from Theorem 5.7 (Van Der Vaart, 2000), it follows that
$$\hat{\beta}_n \stackrel{P}{\rightarrow} \beta_0 ~~\mbox{as}~~n \rightarrow \infty.$$

For (2), we can write (from Theorem-1)
\begin{equation} 
\hat{\Lambda}_{\hat{\beta}_n, F_n}- \Lambda_0=\int_{0}^{T} \frac{E_{F_n} dN(u)}{E_{F_n} \gamma(V) Y(u)e^{\hat{\beta}'_n Z}}-\int_{0}^{T} \frac{E_{F_0} dN(u)}{E_{F_0} \gamma(V) Y(u)e^{{\beta_0}' Z}}.
\end{equation}

We know $\hat{\beta}_n \stackrel{P}{\rightarrow} \beta_0$ and $F_n \stackrel{P}{\rightarrow} F_0$ as $n \rightarrow \infty$. Since $(f, F) \rightarrow E_F(f)=\int f dF$ is continuous and $\hat{\Lambda}_{\beta, F}$ is differentiable with respect to $\beta$ and Hadamard differentiable with respect to $F$, so we can write
$$\hat{\Lambda}_{\hat{\beta}_n, F_n}- \Lambda_0=o_p(1) ~~\mbox{as}~~n \rightarrow \infty.$$

So (2) is also proven. Finally we have proved Theorem-2.

\section{Theorem- 3 with proof}
\setcounter{equation}{0}

\textbf{Theorem 3:} Suppose for assumptions (\textbf{A1-A4}), $\hat{\beta}_n \stackrel{P}{\rightarrow} \beta_0$ and $F_n \stackrel{P}{\rightarrow} F_0$ as $n \rightarrow \infty$, then we have
\begin{equation} \label{3.1}
E\bigg[\sqrt{n}\bigg\{\phi_s(T, \delta|\hat{\beta}_n,F_0)-\phi_s(T, \delta|\beta_0,F_0) \bigg\}\bigg]=-E\bigg[\phi_s(T, \delta |\beta_0,F_0)\phi'_s(T, \delta |\beta_0,F_0)\bigg] \bigg\{\sqrt{n}(\hat{\beta}_n- \beta_0) \bigg\}+ o_p(1),
\end{equation}
and 
\begin{equation} \label{3.2}
\begin{aligned}
E\bigg[\sqrt{n}\bigg\{\phi_s(T, \delta|\hat{\beta}_n,F_n)-\phi_s(T, \delta|\hat{\beta}_n,F_0) \bigg\}\bigg]
={} & -E\bigg[\phi_s(T, \delta|\beta_0,F_0)B(T, \delta|\beta_0,F_0)\bigg] \bigg\{\sqrt{n}(F_n- F_0) \bigg\}\\
& +o_p \big(1+\sqrt{n}(\hat{\beta}_n- \beta_0)\big).
\end{aligned}
\end{equation}
\textbf{Proof:} Based on Lemma-1, we know $P(T, \delta|\beta_0,F_0)> \delta >0$ for some positive constant $\delta > 0$. So by the differentiability of $P(T, \delta|\beta,F)$ with respect to $\beta$ and $F$, we have
\begin{equation} \label{3.3}
\frac{\sqrt{n}\big\{P(T, \delta|\hat{\beta}_n,F_0)-P(T, \delta|\beta_0,F_0) \big\}}{P(T,\delta|\beta_0,F_0)}=\phi_s(T,\delta|\beta_0,F_0)\big\{\sqrt{n}(\hat{\beta}_n- \beta_0)\big\}+ o_p(1),
\end{equation}
\begin{equation} \label{3.4}
\frac{\sqrt{n} \big\{P(T, \delta|\hat{\beta}_n,F_n)-P(T, \delta|\hat{\beta}_n,F_0) \big\}}{P(T, \delta|\beta_0,F_0)}=B(T, \delta|\beta_0,F_0)\big\{\sqrt{n}(F_n- F_0)\big\}+ o_p(1).
\end{equation}
In Lemma-1, we showed the class of functions $\big\{ \phi_s(T, \delta|\beta, F): \beta \in \Theta, F \in \zeta_\rho \big\}$ is uniformly bounded. Similarly, we can show the class of functions $\big\{ B(T, \delta|\beta, F): \beta \in \Theta, F \in \zeta_\rho \big\}$ is uniformly bounded. From these results, it follows that there is a $P_0$-square integrable function, such that
\begin{equation} \label{3.5}
\frac{P(T, \delta|\beta',F')-P(T, \delta|\beta,F)}{P(T, \delta|\beta,F)} \leq M' \big(||\beta'-\beta||+||F'-F|| \big),
\end{equation}
where $M'$ is a $P_0$-square integrable function $\forall \beta, \forall \beta' \in \Theta$ and $\forall F, \forall F' \in \zeta_\rho$.

First we start with (\ref{3.1}), for each $n$, the equality
\begin{equation*}
\begin{aligned}
0%={} & E[\frac{1}{t}\{S(x;\theta_t,F_0)-S(x;\theta_0,F_0) \}]\\
={} & \sqrt{n} \bigg\{\int \phi_s(T, \delta|\hat{\beta}_n,F_0)P(T, \delta|\hat{\beta}_n,F_0)dF-\int \phi_s(T, \delta|\beta_0,F_0)P(T, \delta|\beta_0,F_0)dF\bigg\}\\
={} & \sqrt{n} \bigg\{\int \phi_s(T,\delta|\hat{\beta}_n,F_0)P(T, \delta|\beta_0,F_0)dF-\int \phi_s(T, \delta|\beta_0,F_0)P(T, \delta|\beta_0,F_0)dF\\
& +\int \phi_s(T, \delta|\hat{\beta}_n,F_0)P(T, \delta|\hat{\beta}_n,F_0)dF-\int \phi_s(T, \delta|\hat{\beta}_n,F_0)P(T, \delta|\beta_0,F_0)dF \bigg\},\\
%={} & \int \frac{1}{t} \big \{ S(x;\theta_t,F_0)- S(x;\theta_0,F_0) \big\} p(x;\theta_0,F_0)dx\\ & + \int S(x;\theta_t,F_0) \frac{1}{t} \big \{ p(x;\theta_t,F_0)- p(x;\theta_0,F_0) \big\}dx
\end{aligned}
\end{equation*}
holds and we can express the above equation as
\begin{equation} \label{3.6}
\small
\int \sqrt{n} \bigg \{ \phi_s(T, \delta|\hat{\beta}_n,F_0)- \phi_s(T, \delta|\beta_0,F_0) \bigg\} P(T, \delta|\beta_0,F_0)dF=-\int \phi_s(T, \delta|\hat{\beta}_n,F_0) \sqrt{n} \bigg \{ P(T, \delta|\hat{\beta}_n,F_0)- P(T, \delta|\beta_0,F_0) \bigg\}dF.
\end{equation}

By the dominated convergence theorem with (\ref{3.3}), the right hand side of (\ref{3.6}) can be expressed as, when $n \rightarrow \infty$
\begin{eqnarray} \label{3.7}
\lefteqn {-\int \phi_s(T, \delta|\hat{\beta}_n,F_0) \sqrt{n} \bigg \{ P(T, \delta|\hat{\beta}_n,F_0)- P(T, \delta|\beta_0,F_0) \bigg\}dF} \nonumber \\
&=& -\int \phi_s(T, \delta|\hat{\beta}_n,F_0) \frac{\sqrt{n} \big \{ P(T, \delta|\hat{\beta}_n,F_0)- P(T, \delta|\beta_0,F_0) \big\}}{P(T, \delta|\beta_0,F_0)}P(T, \delta|\beta_0,F_0)dF \nonumber \\
&=& -\int \phi_s(T, \delta|\beta_0,F_0)\phi'_s(T, \delta|\beta_0,F_0)P(T, \delta|\beta_0,F_0)\big\{\sqrt{n}(\hat{\beta}_n- \beta_0)\big\}+ o_p(1)\nonumber \\
&=& -E \bigg[\phi_s(T, \delta|\beta_0,F_0)\phi'_s(T, \delta|\beta_0,F_0)\bigg]\bigg\{\sqrt{n}(\hat{\beta}_n- \beta_0)\bigg\}+ o_p(1).
\end{eqnarray}

So from (\ref{3.6}) and (\ref{3.7}), we can write
\begin{equation*}
\small
\begin{aligned}
\int \sqrt{n} \bigg \{ \phi_s(T, \delta|\hat{\beta}_n,F_0)- \phi_s(T, \delta|\beta_0,F_0) \bigg\} P(T, \delta|\beta_0,F_0)dF %={} & -\int S_s(T, \delta|\theta_0,F_0)S'_s(T, \delta|\theta_0,F_0)P(T, \delta|\theta_0,F_0)\big\{\sqrt{n}(\hat{\theta}_n- \theta_0)\big\}+ o(1)\\
={} & -E \bigg[\phi_s(T, \delta|\beta_0,F_0)\phi'_s(T, \delta|\beta_0,F_0)\bigg]\bigg\{\sqrt{n}(\hat{\beta}_n- \beta_0)\bigg\}+ o_p(1).
\end{aligned}
\end{equation*}

So (\ref{3.1}) is proven. Now we prove (\ref{3.2}) by following the similar idea of proving (\ref{3.1}). For each $n$, the following equation holds
\begin{equation*}
\begin{aligned}
0 %={} & E[\frac{1}{t}\{S(x;\theta_t,F_t)-S(x;\theta_t,F_0) \}]\\
={} & \sqrt{n} \bigg\{\int \phi_s(T, \delta|\hat{\beta}_n,F_n)P(T, \delta|\hat{\beta}_n,F_n)dF-\int \phi_s(T, \delta|\hat{\beta}_n,F_0)P(T, \delta|\hat{\beta}_n,F_0)dF\bigg\}\\
={} & \sqrt{n} \bigg\{\int \phi_s(T, \delta|\hat{\beta}_n,F_n)P(T, \delta|\hat{\beta}_n,F_n)dF-\int \phi_s(T, \delta|\hat{\beta}_n,F_0)P(T, \delta|\hat{\beta}_n,F_n)dF\\
& +\int \phi_s(T, \delta|\hat{\beta}_n,F_0)P(T, \delta|\hat{\beta}_n,F_n)dF-\int \phi_s(T, \delta|\hat{\beta}_n,F_0)P(T, \delta|\hat{\beta}_n,F_0)dF \bigg\}.
% ={} & \int \frac{1}{t} \big \{ S(x;\theta_t,F_t)- S(x;\theta_t,F_0) \big\} p(x;\theta_t,F_t)dx\\ & + \int S(x;\theta_t,F_0) \frac{1}{t} \big \{ p(x;\theta_t,F_t)- p(x;\theta_t,F_0) \big\}dx
\end{aligned}
\end{equation*}

We can express the above equation as
\begin{equation} \label{3.8}
\small
\int \sqrt{n} \bigg \{ \phi_s(T, \delta|\hat{\beta}_n,F_n)- \phi_s(T, \delta|\hat{\beta}_n,F_0) \bigg\} P(T, \delta|\hat{\beta}_n,F_n)dF=-\int \phi_s(T, \delta|\hat{\beta}_n,F_0) \sqrt{n} \bigg \{ P(T, \delta|\hat{\beta}_n,F_n)- P(T, \delta|\hat{\beta}_n,F_0) \bigg\}dF.
\end{equation}

By using the dominated convergence theorem with (\ref{3.5}) and Lemma-2, when $n \rightarrow \infty$, the left hand side of (\ref{3.8}) can be derived as 
\begin{eqnarray} \label{3.9}
\small
\lefteqn {\bigg|\bigg| \int \sqrt{n} \bigg \{ \phi_s(T, \delta|\hat{\beta}_n,F_n)- \phi_s(T, \delta|\hat{\beta}_n,F_0) \bigg\} P(T, \delta|\hat{\beta}_n,F_n)dF-\int \sqrt{n} \bigg \{ \phi_s(T, \delta|\hat{\beta}_n,F_n)- \phi_s(T, \delta|\hat{\beta}_n,F_0) \bigg\} P(T, \delta|\beta_0,F_0)dF\bigg|\bigg|}\nonumber \\
& = & \bigg|\bigg| \int \bigg \{ \phi_s(T, \delta|\hat{\beta}_n,F_n)- \phi_s(T, \delta|\hat{\beta}_n,F_0) \bigg\} \sqrt{n} \bigg\{P(T, \delta|\hat{\beta}_n,F_n)- P(T, \delta|\beta_0,F_0)\bigg\}dF \bigg|\bigg| \nonumber \\
& = & \bigg|\bigg| \int \bigg \{ \phi_s(T, \delta|\hat{\beta}_n,F_n)- \phi_s(T, \delta|\hat{\beta}_n,F_0) \bigg\}  \frac{\sqrt{n} \big\{P(T, \delta|\hat{\beta}_n,F_n)- P(T, \delta|\beta_0,F_0)\big\}}{P(T, \delta|\beta_0,F_0)}P(T, \delta|\beta_0,F_0) dF \bigg|\bigg| \nonumber \\
& \leq & \bigg| \int  \bigg \{M'' \bigg(\big|\big|F_n-F_0 \big|\big|+ \big|\big| \hat{\Lambda}_{\hat{\beta}_n, F_n}- \hat{\Lambda}_{\hat{\beta}_n, F_0} \big|\big| \bigg)    \bigg  \} \bigg\{\sqrt{n}~M' \bigg(\big|\big|\hat{\beta}_n- \beta_0 \big|\big|+ \big|\big|F_n-F_0 \big|\big|\bigg) \bigg\} P(T, \delta|\beta_0,F_0) dF \bigg| \nonumber \\
& = & \bigg\{\bigg| \int M'' M' P(T, \delta|\beta_0,F_0) dF  \bigg| \bigg\} \times \bigg\{\sqrt{n}\bigg(\big|\big|\hat{\beta}_n- \beta_0 \big|\big|+ \big|\big|F_n-F_0 \big|\big|\bigg) \bigg(\big|\big|F_n-F_0 \big|\big|+ \big|\big|\hat{\Lambda}_{\hat{\beta}_n, F_n}-\hat{\Lambda}_{\hat{\beta}_n, F_0} \big|\big|\bigg)\bigg\} \nonumber \\
& = & O\bigg\{\sqrt{n}\bigg(\big|\big|\hat{\beta}_n- \beta_0 \big|\big|+ \big|\big|F_n- F_0 \big|\big|\bigg)\bigg(\big|\big|F_n-F_0 \big|\big|+ \big|\big|\hat{\Lambda}_{\hat{\beta}_n, F_n}-\hat{\Lambda}_{\hat{\beta}_n, F_0} \big|\big|\bigg) \bigg\} \nonumber \\
%& = & O\bigg\{\sqrt{n}\big(\big|\big|\hat{\beta}_n- \beta_0 \big|\big|\big)\bigg(\big|\big|F_n-F_0 \big|\big|+ \big|\big|\hat{\Lambda}_{\hat{\beta}_n, F_n}-\hat{\Lambda}_{\hat{\beta}_n, F_0} \big|\big|\bigg) \bigg\} +\bigg\{\sqrt{n}\big( \big|\big|F_n- F_0 \big|\big|\big)\bigg(\big|\big|F_n-F_0 \big|\big|+ \big|\big|\hat{\Lambda}_{\hat{\beta}_n, F_n}-\hat{\Lambda}_{\hat{\beta}_n, F_0} \big|\big|\bigg) \bigg\} \nonumber \\
& = & \sqrt{n}\big(\hat{\beta}_n- \beta_0 \big) O \big(o_p(1) \big)+O \big(O_p(1).o_p(1) \big) \nonumber \\
& = & \sqrt{n}\big(\hat{\beta}_n- \beta_0 \big).o_p(1)+o_p(1) \nonumber \\
& = & o_p \big(1+\sqrt{n}(\hat{\beta}_n- \beta_0)\big),
\end{eqnarray}
where we used $F_n-F_0=o_p(1)$ from assumption (\textbf{A3}),  $\hat{\beta}_n-\beta_0=o_p(1)$ and $\hat{\Lambda}_{\hat{\beta}_n, F_n}-\hat{\Lambda}_{{\beta}_0, F_0}=o_p(1)$ from Theorem-2.

By the dominated convergence theorem with (\ref{3.4}), the right hand side of (\ref{3.8}) can be written as, when $n \rightarrow \infty$
\begin{eqnarray} \label{3.10}
\lefteqn { -\int \phi_s(T, \delta|\hat{\beta}_n,F_0) \sqrt{n} \bigg \{ P(T, \delta|\hat{\beta}_n,F_n)- P(T, \delta|\hat{\beta}_n,F_0) \bigg\}dF} \nonumber \\
&= & -\int \phi_s(T, \delta|\hat{\beta}_n,F_0)  \frac{\sqrt{n}  \big\{P(T, \delta|\hat{\beta}_n,F_n)- P(T, \delta|\hat{\beta}_n,F_0) \big\}}{P(T, \delta|\beta_0,F_0)}P(T, \delta|\beta_0,F_0)dF \nonumber \\
& = & -\int \phi_s(T, \delta|\beta_0,F_0)B(T, \delta|\beta_0,F_0) P(T, \delta|\beta_0,F_0) \sqrt{n}(F_n- F_0)dF+ o_p(1)\nonumber \\
& = & -E\bigg[\phi_s(T, \delta |\beta_0,F_0)B(T, \delta |\beta_0,F_0)\bigg]\sqrt{n}(F_n- F_0)+ o_p(1).
\end{eqnarray}

So by combining (\ref{3.9}) and (\ref{3.10}), the equality (\ref{3.8}) is equivalent to
\begin{eqnarray}
\lefteqn { \int \sqrt{n} \bigg \{ \phi_s(T, \delta|\hat{\beta}_n,F_n)- \phi_s(T, \delta|\hat{\beta}_n,F_0) \bigg\} P(T, \delta|{\beta}_0,F_0)dF} \nonumber \\
% & = & -\int S_s(T, \delta|\beta_0,F_0)B(T, \delta|\beta_0,F_0)\sqrt{n}(F_n- F_0)P(T, \delta|\beta_0,F_0)dF +o_p \big(1+\sqrt{n}(\hat{\beta}_n- \beta_0)\big) \nonumber \\
& = & -E\bigg[\phi_s(T, \delta |\beta_0,F_0)B(T, \delta |\beta_0,F_0)\bigg]\sqrt{n}(F_n- F_0) +o_p \big(1+\sqrt{n}(\hat{\beta}_n- \beta_0)\big).
\end{eqnarray}

So equation (\ref{3.2}) is also proven. Hence, we proved Theorem-3.

\section{Theorem- 4 with proof}
\setcounter{equation}{0}
\textbf{Theorem 4:} If the assumptions $(\textbf{A1-A4})$ are satisfied, then a consistent estimator $\hat{\beta}_n$ to the estimating equation
\begin{equation}
\sum_{i=1}^{n}\phi_s(T_i, \delta_i|\hat{\beta}_n, F_n)=0,
\end{equation}
is an asymptotically linear estimator for $\beta_0$ (Hirose, 2011a) with the efficient influence function $(I_s^*)^{-1}\phi_s(T, \delta|\beta_0, F_0)$, so that
$$\sqrt{n} (\hat{\beta}_n-\beta_0)=\frac{1}{\sqrt{n}}\sum_{i=1}^{n}(I_s^*)^{-1}\phi_s(T_i, \delta_i|\beta_0, F_0)+o_p(1) \stackrel{D}{\longrightarrow}N\{0,(I_s^*)^{-1}\},$$
where $N\{0,(I_s^*)^{-1}\}$ is a normal distribution with mean zero and variance $(I_s^*)^{-1}$. So the estimator $\hat{\beta}_n$ is efficient.

In addition, we know that $\phi_l(V|b)$ is the score function of logistic regression (which is a parametric model), then a consistent estimator $\hat{b}_n$ to the estimating equation
\begin{equation*}
\sum_{i=1}^{n}\phi_l(V_i|\hat{b}_n)=0,
\end{equation*}
is an asymptotically linear estimator for $b_0$ with the influence function $(I_l)^{-1}\phi_l(V|b_0)$, so that
$$\sqrt{n} (\hat{b}_n-b_0)=\frac{1}{\sqrt{n}}\sum_{i=1}^{n}(I_l)^{-1}\phi_l(V_i|b_0)+o_p(1) \stackrel{D}{\longrightarrow}N\{0,(I_l)^{-1}\},$$
where $I_l=E[\phi_l \phi'_l]$ and $N\{0,(I_l)^{-1}\}$ is a normal distribution with mean zero and variance $(I_l)^{-1}$.\\
\textbf{Proof:} Since $\phi_s(T_, \delta| \beta, F)$ is uniformly bounded Donsker (Lemma-1). So by Lemma 19.24 (Van Der Vaart, 2000), we can write
\begin{equation} \label{4.2}
\frac{1}{\sqrt{n}}\sum_{i=1}^{n} \bigg\{\phi_s(T_i, \delta_i|\hat{\beta}_n, F_n)-\phi_s(T_i, \delta_i| \beta_0, F_0)\bigg\}=\sqrt{n}E\bigg[\phi_s(T, \delta|\hat{\beta}_n, F_n)-\phi_s(T, \delta| \beta_0, F_0)\bigg]+o_p(1),
\end{equation}

From (\ref{3.1}), it follows that
\begin{equation} \label{4.3}
\begin{aligned}
\sqrt{n} E\bigg[\phi_s(T, \delta|\hat{\beta}_n,F_0)-\phi_s(T, \delta|\beta_0,F_0)\bigg]={} & -E\bigg[\phi_s(T, \delta|\beta_0,F_0)\phi'_s(T, \delta|\beta_0,F_0)\bigg]\sqrt{n}(\hat{\beta}_n- \beta_0)+ o_p(1)\\
={} & - I_s^{*}\sqrt{n}(\hat{\beta}_n- \beta_0)+ o_p(1).
\end{aligned}
\end{equation}

From (\ref{3.2}),  it follows that
\begin{equation} \label{4.4}
\begin{aligned}
\sqrt{n} E\bigg[\phi_s(T, \delta|\hat{\beta}_n,F_n)-\phi_s(T, \delta|\hat{\beta}_n,F_0) \bigg]
=& -E\bigg[\phi_s(T, \delta|\beta_0,F_0)B(T, \delta|\beta_0,F_0)\bigg]\sqrt{n}(F_n- F_0)\\
& +o_p \big(1+\sqrt{n}(\hat{\beta}_n- \beta_0)\big).
%\\ & +o(1+||F_n-F_0||+||\hat{\eta}_{\hat{\theta}_n, F_n}-\eta_0||)
\end{aligned}
\end{equation}	

%Finally express the above equation as
%\begin{equation}
%\sqrt{n} E\bigg[S_s(T, \delta|\hat{\beta}_n,F_n)-S_s(T, \delta|\hat{\beta}_n,F_0) \bigg]=o_p \big(1+\sqrt{n}(\hat{\beta}_n- \beta_0)\big).
%\end{equation}

Since $B(T, \delta|\beta_0,F_0)$ is in the nuisance tangent space and $\phi_s(T, \delta|\beta_0,F_0)$ is the efficient score function, so we can consider
\begin{equation} \label{4.5}
E\big[\phi_s(T, \delta|\beta_0,F_0)B(T, \delta|\beta_0,F_0)\big]=0.
\end{equation}
%and \begin{equation*} o(1+||F_n-F_0||+||\hat{\eta}_{\hat{\theta}_n, F_n}-\eta_0||)=o_p(1) \end{equation*}

Now using (\ref{4.3}), (\ref{4.4}) and (\ref{4.5}), the right hand side of (\ref{4.2}) can be expressed as
\begin{equation} \label{4.6}
\small
\begin{aligned}
\sqrt{n}E\bigg[\phi_s(T, \delta|\hat{\beta}_n, F_n)-\phi_s(T, \delta| \beta_0, F_0)\bigg]={} & \sqrt{n} E\bigg[\phi_s(T, \delta|\hat{\beta}_n,F_0)-\phi_s(T, \delta|\beta_0,F_0) \bigg] +\sqrt{n} E\bigg[\phi_s(T, \delta|\hat{\beta}_n,F_n)-\phi_s(T, \delta|\hat{\beta}_n,F_0) \bigg]\\
={} & - I_s^{*}\sqrt{n}(\hat{\beta}_n- \beta_0)+o_p \big(1+\sqrt{n}(\hat{\beta}_n- \beta_0)\big).
\end{aligned}
\end{equation}

We know that $\frac{1}{\sqrt{n}}\sum_{i=1}^{n}\phi_s(T_i, \delta_i|\hat{\beta}_n, F_n)=0$, so using (\ref{4.6}), the equation (\ref{4.2}) can be written as 
\begin{equation} \label{4.7}
\begin{aligned}
\big(I_s^*+ o_p(1) \big) \sqrt{n}(\hat{\beta}_n- \beta_0)={} & \frac{1}{\sqrt{n}}\sum_{i=1}^{n} \phi_s(T_i, \delta_i| \beta_0, F_0)+ o_p(1).\\
%\sqrt{n}(\hat{\beta}_n- \beta_0)={} & \frac{1}{\sqrt{n}} \sum_{i=1}^{n} I_s^{*-1} S_s(T_i, \delta_i| \beta_0, F_0)+o_p(1).
\end{aligned}
\end{equation}

By Central Limit Theorem (CLT), we can write $\frac{1}{\sqrt{n}} \sum_{i=1}^{n}  \phi_s(T_i, \delta_i| \beta_0, F_0)+o_p(1)=O_p(1)$. 
%since we know that $O_p(1)+o_p(1)=O_p(1)$ where $\frac{1}{\sqrt{n}} \sum_{i=1}^{n}  S_s(T_i, \delta_i| \beta_0, F_0)=O_p(1)$. Similarly, it follows that $\big(I_s^*+ o_p(1) \big)=O_p(1)$. 
Since $I_s^*$ is invertible, we have $\big(I_s^*+ o_p(1) \big)^{-1}=O_p(1)$.

So from (\ref{4.7}) we can write $\sqrt{n}(\hat{\beta}_n- \beta_0)=\big(I_s^*+ o_p(1) \big)^{-1} O_p(1)=O_p(1)$. 
%since we know $O_p(1) O_p(1)=O_p(1)$. Hence from (3.26), it follows that $o_p \big(1+\sqrt{n}(\hat{\beta}_n- \beta_0)\big)=o_p(1)$, since $o_p \big(1+ O_p(1) \big)=o_p(1)$.

%We know that $O_p(1) O_p(1)=O_p(1)$, so from (3.27) we can write $\sqrt{n}(\hat{\beta}_n- \beta_0)=O_p(1)$. Hence from (3.26), it follows that $o_p \big(1+\sqrt{n}(\hat{\beta}_n- \beta_0)\big)=o_p(1)$, since $o_p \big(1+ O_p(1) \big)=o_p(1)$.
Finally we can express (\ref{4.2}) as
\begin{equation*}
\begin{aligned}
%\big(I_s^*+ o_p(1) \big) \sqrt{n}(\hat{\beta}_n- \beta_0)={} & \frac{1}{\sqrt{n}}\sum_{i=1}^{n} S_s(T_i, \delta_i| \beta_0, F_0)+ o_p(1)\\
\sqrt{n}(\hat{\beta}_n- \beta_0)={} & \frac{1}{\sqrt{n}} \sum_{i=1}^{n} I_s^{*-1} \phi_s(T_i, \delta_i| \beta_0, F_0)+o_p(1).
\end{aligned}
\end{equation*}

It follows that the large sample distribution of the estimator $\hat{\beta}_n$ can be expressed as
\begin{equation*}
\sqrt{n}(\hat{\beta}_n- \beta_0) \stackrel{D}{\longrightarrow}N \big \{0,(I_s^*)^{-1} \big \},
\end{equation*}
where $I_s^*=E[\phi^*_{\beta}\phi^{*'}_{\beta}]$ is the efficient information ($\phi^*_{\beta}$ is the efficient score function defined in Theorem- 1). %The logistic part is a parametric model and it is differentiable with respect to the parameter, $b$ (we omit the calculation).
\section{Efficient Score Function for Cox PH Cure Model using Projection Theory}
\setcounter{equation}{0}
To get the efficient score function using the projection theory, we assume the parameters $(\beta, \Lambda)$ are evaluated at the true values $\beta_0$, $\Lambda_0$ and omit subscript ``0" for brevity.

The log-likelihood function of the survival part for one observation can be written as
\begin{equation*}
\log P(T, \delta|\beta, \Lambda)= \gamma(V) \bigg\{ \delta \big( \log \lambda(t)+\beta' Z     \big)   -e^{\beta' Z} \Lambda(t) \bigg\}.
\end{equation*}
\subsection*{Score Function for $\beta$}
\begin{equation*}
\begin{aligned}
\phi_\beta(T, \delta|\beta, \Lambda)={} & \frac{\partial}{\partial \beta} \log P(T, \delta|\beta, \Lambda)=\gamma(V) \bigg\{ Z \big( \delta -e^{\beta' Z} \Lambda(t) \big) \bigg\}.
%\\={} & \gamma(V) \frac{\partial}{\partial \theta} \bigg\{ \delta \big( \log \lambda(t)+\beta' Z     \big)   -e^{\beta' Z} \Lambda(t) \bigg\}\\
%={} & \gamma(V) \bigg\{ Z \big( \delta -e^{\beta' Z} \Lambda(t) \big) \bigg\}.
\end{aligned}
\end{equation*}
\subsection*{Score Operator for $\Lambda$}
Let us take a measurable function which is bounded such as
$g : [0, \tau ] \rightarrow R$, where $g$ is defined in the interval $[0, \tau ]$ because $\Lambda$ is also restricted within this interval.
The path can be defined as $$d\Lambda_s=(1+sg)d\Lambda. $$

The corresponding path for the baseline hazard function is
$$\lambda_s(t)=\frac{d\Lambda_s}{dt}=(1+sg)\frac{d\Lambda}{dt}=(1+sg)\lambda(t) .$$

%The log likelihood function can be expressed as
%\begin{equation*}
%\log P(T, \delta|\beta, \Lambda_s)= \gamma(V) \bigg\{ \delta \big( \log \lambda_s(t)+\beta' Z     \big)   -e^{\beta' Z} \Lambda_s(t) \bigg\}.
%\end{equation*}

The derivative of the log-likelihood function with respect to $s$ can be expressed as
%\begin{equation*}
%\begin{aligned}
%\frac{\partial}{\partial s} \log P(T, \delta|\beta, \Lambda_s) 
%={} & \gamma(V) \bigg\{ \delta \frac{\partial}{\partial s}\log \lambda_s(t) - e^{\beta'Z}\frac{\partial}{\partial s}\Lambda_s(t) \bigg\}=\gamma(V) \bigg\{ \frac{\delta f(t)}{(1+sf)}-e^{\beta'Z}\int_{0}^{t} f(u) d\Lambda(u) \bigg\}.\\
%={} & \gamma(V) \bigg\{ \delta \frac{\partial}{\partial s} \log [(1+sf)\lambda(t)]- e^{\beta'Z}  \frac{\partial}{\partial s} \int_{0}^{t} (1+sf)d\Lambda(u) \bigg\}\\
%={} & \delta \frac{1}{(1+sa)\lambda(t)}a\lambda(t)-e^{\beta'Z}\int_{0}^{t} a(u) d\Lambda(u)\\
%={} & \gamma(V) \bigg\{ \frac{\delta f(t)}{(1+sf)}-e^{\beta'Z}\int_{0}^{t} f(u) d\Lambda(u) \bigg\}.
%\end{aligned}
%\end{equation*}
%When $s=0$, the derivative yields $\gamma(V) \bigg\{\delta f(t)-e^{\beta'Z}\int_{0}^{t} f(u) d\Lambda(u)\bigg\}$, which is the score operator of $\Lambda$ and can be viewed as 
$$ B_\Lambda(T, \delta|\beta, \Lambda) g=\frac{\partial}{\partial s} \bigg|_{s=0} \log P(T, \delta|\beta, \Lambda_s)=\gamma(V) \bigg\{\delta g(t)-e^{\beta'Z}\int_{0}^{t} g(u) d\Lambda(u)\bigg\}.$$
\subsection*{Information Operator $B^*_\Lambda B_\Lambda$ and its Inverse $\big(B^*_\Lambda B_\Lambda\big)^{-1}$}
Let us start with the information operator $B^*_\Lambda B_\Lambda$ and take two arbitrary functions $f$ and $g$. By definition of the adjoint, we can write
\begin{equation} \label{5.1}
\langle B^*_\Lambda B_\Lambda f,g\rangle_{L_2(\Lambda)}= \langle B_\Lambda f, B_\Lambda g\rangle_{L_2(P)}.
\end{equation}

The path defined by $d\Lambda_{r,s}=(1+rf+sg+rsfg)d\Lambda$ is positive for small $r$ and $s$. It can be written as $d\Lambda_{r,s}=(1+rf)(1+sg)d\Lambda$.
%So we can write $d\Lambda_{r,0}=(1+rf)d\Lambda$ and $d\Lambda_{0,s}=(1+sg)d\Lambda.$
The corresponding path for the baseline hazard function is
$$\lambda_{r,s}(t)=\frac{d\Lambda_{r,s}}{dt}=(1+rf+sg+rsfg)\frac{d\Lambda}{dt}=(1+rf+sg+rsfg)\lambda(t) .$$
%Then the log likelihood function for the submodel will be
%\begin{equation*}
%\log P(T, \delta|\beta, \Lambda_{r,s})= \gamma(V) \bigg\{ \delta \big( \log \lambda_{r,s}(t)+\beta' Z     \big)   -e^{\beta' Z} \Lambda_{r,s}(t) \bigg\}.
%\end{equation*}
%The derivative of the log likelihood with respect to $r$ can be written as
%\begin{equation}
%\begin{aligned}
%\frac{\partial}{\partial r} \log P(T, \delta|\beta, \Lambda_{r,s})
%={} & \gamma(V) \bigg\{ \delta \frac{\partial}{\partial r}\log \lambda_{r,s}(t) - e^{\beta'Z}\frac{\partial}{\partial r}\Lambda_{r,s}(\xi) \bigg\}=\gamma(V) \bigg\{ \frac{\delta f(t)}{1+rf(t)}-e^{\beta'Z} \int_{0}^{t}f(\xi) d\Lambda_{0,s}(\xi)\bigg\}~.\\
%={} & \gamma(V) \bigg\{ \delta \frac{\partial}{\partial r} \log \bigg[(1+rf+sg+rsfg)\lambda(t)\bigg]\\
%&- e^{\beta'Z} \frac{\partial}{\partial r}\int_{0}^{t} (1+rf+sg+rsfg)d\Lambda(\xi) \bigg\}\\
%={} & \gamma(V) \bigg\{ \delta \frac{(f+sfg)\lambda(t)}{(1+rf+sg+rsfg)\lambda(t)}-e^{\beta'Z} \int_{0}^{t}f(1+sg) d\Lambda(\xi) \bigg\}\\
%={} & \gamma(V) \bigg\{ \frac{\delta f(t)}{1+rf(t)}-e^{\beta'Z} \int_{0}^{t}f(\xi) d\Lambda_{0,s}(\xi)\bigg\}~.~~~~~~~~~~~~~~~~~~~~~~~~~~~~~~~~~(*)
%\end{aligned}
%\end{equation}

Now we can write
%When $(r,s)=(0,0)$, the derivative yields  
\begin{equation} \label{5.2}
\frac{\partial}{\partial r}\bigg|_{(r,s)=(0,0)}\log P(T, \delta|\beta, \Lambda_{r,s})%=\gamma(V) \bigg\{\delta f(t)-e^{\beta'Z} \int_{0}^{t}f(\xi) d\Lambda(\xi) \bigg\}
=B_\Lambda f,
\end{equation}

and
\begin{equation} \label{5.3}
\frac{\partial}{\partial s}\bigg|_{(r,s)=(0,0)}\log P(T, \delta|\beta, \Lambda_{r,s})=B_\Lambda g.
\end{equation}

Using (\ref{5.2}) and (\ref{5.3}) we can write 
\begin{equation} \label{5.4}
\begin{aligned}
\langle B_\Lambda f, B_\Lambda g\rangle_{L_2(P)}={} & E\bigg\{ (B_\Lambda f) (B_\Lambda g)\bigg\}\\
={} & -E\bigg\{\frac{\partial^2}{\partial r \partial s}\bigg|_{(r,s)=(0,0)}\log P(T, \delta|\beta, \Lambda_{r,s})\bigg\}\\
={} & E \bigg\{ \gamma(V) e^{\beta'Z} \int_{0}^{t}f(\xi)g(\xi) d\Lambda(\xi) \bigg\}.
%={} &  E\bigg\{\frac{\partial}{\partial r}\bigg|_{(0,0)}\log P(T, \delta|\theta, \Lambda_{r,s})\frac{\partial}{\partial s}\bigg|_{(0,0)}\log P(T, \delta|\theta, \Lambda_{r,s})\bigg\}\\
%={} & -E\bigg\{\frac{\partial^2}{\partial r \partial s}\bigg|_{(0,0)}\log P(T, \delta|\theta, \Lambda_{r,s})\bigg\}.
\end{aligned}
\end{equation}

%From equation (4.29) we can observe that
%\begin{equation}
%-E\bigg\{\frac{\partial^2}{\partial r \partial s}\bigg|_{(0,0)}\log P(T, \delta|\beta, \Lambda_{r,s})\bigg\}=E \bigg\{ \gamma(V) e^{\beta'Z} \int_{0}^{t}f(\xi)g(\xi) d\Lambda(\xi) \bigg\}.
%\end{equation}

Now we manipulate the integral involving the function $\xi$, we deduce
\begin{equation*}
\int_{0}^{t}f(\xi)g(\xi) d\Lambda(\xi)=\int_{0}^{\tau}I(\xi \leq T)f(\xi)g(\xi)d\Lambda(\xi).
\end{equation*}

Indeed, if $\xi > T$, then the contribution will be 0 to the integral. So the last term in equation (\ref{5.4}) can be expressed as
\begin{equation} \label{5.5}
E\bigg\{\gamma(V)e^{\beta'Z} \int_{0}^{t}f(\xi)g(\xi) d\Lambda(\xi) \bigg\}=E\bigg\{\gamma(V)e^{\beta'Z}\int_{0}^{\tau}I(\xi \leq T)f(\xi)g(\xi)d\Lambda(\xi) \bigg\}.
\end{equation}

Using Fubini's theorem, equation (\ref{5.5}) can be written as
\begin{equation}
\begin{aligned}
E\bigg\{\gamma(V)e^{\beta'Z}\int_{0}^{\tau}I(\xi \leq T)f(\xi)g(\xi)d\Lambda(\xi) \bigg\} %\int \bigg[e^{\beta'Z}\int_{0}^{\tau}I(\xi \leq T)a(\xi)b(\xi)d\Lambda(\xi)\bigg]p_{\beta, \Lambda}dP_{\beta, \Lambda}\\
%={} & \int_{0}^{\tau} g(\xi)E\bigg[ \big\{\gamma(V)e^{\beta'Z} I(\xi \leq T)\big\}f(\xi)\bigg]d\Lambda(\xi)\\
={} & \bigg\langle E \big\{\gamma(V)e^{\beta'Z} I(\xi \leq T)f(\xi)\big\},g(\xi) \bigg\rangle_{L_2(\Lambda)}.
\end{aligned}
\end{equation}

From equation (\ref{5.1}) we can write
\begin{equation*}
\bigg\langle B^*_\Lambda B_\Lambda f,g\bigg\rangle_{L_2(\Lambda)}= \bigg\langle E \{\gamma(V)e^{\beta'Z} I(t \leq T)f\},g \bigg\rangle_{L_2(\Lambda)}.
\end{equation*}

So, the information operator is
\begin{equation*}
B^*_\Lambda B_\Lambda f=E \bigg\{\gamma(V)e^{\beta'Z} I(t \leq T)\bigg\}f(t).
\end{equation*}

It follows that the inverse of information operator is
\begin{equation*}
\big(B^*_\Lambda B_\Lambda \big)^{-1} f(t)=\bigg[E \{\gamma(V)e^{\beta'Z} I(t \leq T)\}\bigg]^{-1}f(t).
\end{equation*}

\subsection*{The Action of the Adjoint Score Operator $B^*_\Lambda$ on the Score Function $\phi_\beta$}
Assume the differentiable paths $(r,s) \mapsto P(T,\delta|\beta+ru, \Lambda_s)$ can be exploited with the path $d\Lambda_s=(1+sg)d\Lambda$. Now we can write %where $\Lambda_s$ is well defined by the boundedness of $g$ and the derivative with respect to $s$ exist. 
%The corresponding path for the baseline hazard function can be written as
%$$\lambda_s(t)=\frac{d\Lambda_s}{dt}=(1+sg)\frac{d\Lambda}{dt}=(1+sg)\lambda(t). $$

%Then the log likelihood function can be written as
%\begin{equation*}
%\log P(T,\delta|\beta+ru, \Lambda_s)= \gamma(V) \bigg\{ \delta \big( \log \lambda_{s}(t)+(\beta+ru)' Z     \big)   -e^{(\beta+ru)' Z} \Lambda_{s}(t) \bigg\}.
%\end{equation*}

%The derivative of the log likelihood with respect to $r$ can be expressed as
%\begin{equation*}
%\frac{\partial}{\partial r} \log P(T,\delta|\beta+ru, \Lambda_s)=\gamma(V) \bigg\{  \delta u'Z-u'Ze^{(\beta+ru)'Z} \int_{0}^{t} (1+sg)d\Lambda(\xi) \bigg\}.
%\end{equation*}

%When $(r,s)=(0,0)$, then the score function can be written as
\begin{equation} \label{5.7}
\frac{\partial}{\partial r}\bigg|_{(r,s)=(0,0)} \log P(T,\delta|\beta+ru, \Lambda_s)%\gamma(V) \bigg\{\delta u'Z-u'Ze^{\beta'Z}\Lambda(t) \bigg\}
= u'\phi_\beta.
\end{equation}
and 
%Now the derivative of the log-likelihood with respect to $s$ and evaluating at $(r,s) = (0, 0)$ can be expressed as
\begin{equation} \label{5.8}
\frac{\partial}{\partial s}\bigg|_{(r,s)=(0,0)} \log P(T,\delta|\beta+ru, \Lambda_s)=B_\Lambda g.
\end{equation}

%When $(r,s)=(0,0)$, then the score function can be written as
%\begin{equation}
%\gamma(V) \bigg\{ \delta g(t)-e^{\beta'Z} \int_{0}^{t}g(\xi) d\Lambda(\xi) \bigg\}=B_\Lambda g.
%\end{equation}

Using equation (\ref{5.7}) and (\ref{5.8}) we can write
\begin{equation} \label{5.9}
\begin{aligned}
\bigg \langle u'\phi_\beta, B_\Lambda g\bigg \rangle
={} & E\bigg\{ (u'\phi_\beta) (B_\Lambda g)\bigg\}\\
={} & -E\bigg\{\frac{\partial^2}{\partial r \partial s}\bigg|_{(r,s)=(0,0)}\log P(T, \delta|\beta+ru, \Lambda_{s})\bigg\}\\
={} & u' E	\bigg\{\gamma(V) Ze^{\beta'Z} \int_{0}^{t} g(\xi) d\Lambda(\xi)  \bigg\}.
%={} &  E\bigg\{\frac{\partial}{\partial r}|_{(0,0)}\log P(T,\delta|\theta+ru, \Lambda_s) \times \frac{\partial}{\partial s}|_{(0,0)}\log P(T,\delta|\theta+ru, \Lambda_s)\bigg\}\\
%={} & -E\bigg\{\frac{\partial^2}{\partial r \partial s}|_{(0,0)}\log P(T,\delta|\theta+ru, \Lambda_s)\bigg\}\\
%={} & -E_{\beta, \Lambda}\bigg[\frac{\partial}{\partial s}|_{(s=0)} \{\frac{\partial}{\partial r}|_{(r=0)}\log p_{\beta+ru, \Lambda_{s}}(X) \}\bigg]\\
%={} & -u'E_{\beta, \Lambda}\bigg\{\frac{\partial}{\partial s}|_{(s=0)}\delta Z-Ze^{\beta'Z}  \int_{0}^{t} (1+sg)d\Lambda(\xi) \bigg \}\\
%={} & -E	\bigg\{-u'\gamma(V)Ze^{\beta'Z} \int_{0}^{t}g(\xi)d\Lambda(\xi)  \bigg\}\\
%={} & u' E	\bigg\{\gamma(V) Ze^{\beta'Z} \int_{0}^{t} g(\xi) d\Lambda(\xi)  \bigg\}.	
\end{aligned}
\end{equation}

Now by manipulating the integral involving the function $\xi$, the equation (\ref{5.9}) can be expressed as% $\xi$, we deduce
%\begin{equation*}
%\int_{0}^{t} g(\xi) d\Lambda(\xi)=\int_{0}^{\tau}I(\xi \leq T)g(\xi) d\Lambda(\xi)
%\end{equation*}
%Indeed, if $\xi > T$, then the contribution will be 0 to the integral. So (4.16) can be expressed as
\begin{equation}
\bigg \langle u'\phi_\beta,B_\Lambda g\bigg \rangle=u' E\bigg\{\gamma(V) Ze^{\beta'Z}\int_{0}^{\tau}I(\xi \leq T)g(\xi) d\Lambda(\xi) \bigg\}.
\end{equation}

Using the Fubini's theorem, we can conclude that
\begin{equation}
\begin{aligned}
u' E\bigg\{\gamma(V) Ze^{\beta'Z}\int_{0}^{\tau}I(\xi \leq T)g(\xi) d\Lambda(\xi) \bigg\}%={} & \int \bigg[Ze^{\beta'Z}I(\xi \leq T)b(\xi) \Lambda(\xi) \bigg]p_{\beta, \Lambda}dP_{\beta, \Lambda}\\
%={} & u' \int_{0}^{\tau}g(\xi)E\bigg[ \gamma(V) Ze^{\beta'Z} I(\xi \leq T)P_{\theta, \Lambda}dP_{\theta, \Lambda}\bigg] d\Lambda(\xi)\\
={} & \bigg \langle u' E \bigg\{\gamma(V) Ze^{\beta'Z} I(\xi \leq T)\bigg\},g(\xi) \bigg \rangle_{L_2(\Lambda)}.\\
%={} & \bigg \langle B^*_{\beta, \Lambda}S_\beta(Z;\beta, \Lambda), b(t)\bigg \rangle
\end{aligned}
\end{equation}

We know that
\begin{equation*}
%u'\bigg \langle B^*_\Lambda S_\beta,g\bigg \rangle= 
\bigg \langle u' B^*_\Lambda \phi_\beta,g\bigg \rangle_{{L_2}(P)}= \bigg \langle u'\phi_\beta, B_\Lambda g\bigg \rangle_{L_2(\Lambda)}.
\end{equation*}

So we can write
$$ B^*_\Lambda \phi_\beta=E\bigg\{\gamma(V)Ze^{\beta'Z} I(t \leq T)\bigg\}.$$
\subsection*{Efficient Score Function $ \phi^*_\beta$:}
Finally the efficient score function can be expressed as
\begin{equation}
\begin{aligned}
\phi^*_\beta %={} & \bigg[I-B_\Lambda \big(B^*_\Lambda B_\Lambda \big)^{-1}B^*_\Lambda \bigg] S_\theta\\
={} & \phi_\beta-B_\Lambda \big(B^*_\Lambda B_\Lambda \big)^{-1}B^*_\Lambda \phi_\beta\\
%={} & S_\beta(Z;\beta, \Lambda)(t)-B_{\beta, \Lambda}(B^*_{\beta, \Lambda}B_{\beta, \Lambda})^{-1}\bigg[E_{\beta, \Lambda} \{Ze^{\beta'Z} I(t \leq T)\}\bigg]\\
={} & \gamma(V) \bigg\{ \delta Z-Ze^{\beta'Z} \Lambda(T)- \bigg[\delta-e^{\beta'Z}\int_{0}^{T} d\Lambda (u)\bigg] \frac{E [ \gamma(V) Ze^{\beta'Z} I(t \leq T)]}{E [\gamma(V) e^{\beta'Z} I(t \leq T)]} \bigg\}\\
%={} & \gamma(V) \bigg\{ \delta Z-Ze^{\beta'Z} \Lambda(t)-\bigg[\delta-e^{\beta'Z}\int_{0}^{t} d\Lambda (s)\bigg] \frac{M_1(u)}{M_0(u)} \bigg\}\\
={} & \gamma(V) \bigg\{ \delta \bigg[Z-\frac{M_1(T)}{M_0(T)}\bigg] -e^{\beta'Z}\int_{0}^{T} \bigg[Z-\frac{M_1(u)}{M_0(u)}\bigg]d\Lambda (u) \bigg\},
\end{aligned}
\end{equation}
where $M_0(T)$ and $M_1(T)$ were defined in the proof of Theorem 1.

\end{document}